\documentclass[useAMS,usenatbib]{mn2e}
 \usepackage{graphicx}
 \usepackage[cp1251]{inputenc}
 \usepackage[T2A]{fontenc}

 \title[Analysis of galaxy kinematics based on Cepheids]
 {Analysis of galaxy kinematics based on Cepheids from the Gaia DR2 Catalogue}

 \author[V. V. Bobylev, A. T. Bajkova, A. S. Rastorguev, and M. V. Zabolotskikh]
        {V. V. Bobylev$^1$\thanks{E-mail: vbobylev@gaoran.ru}, A. T. Bajkova$^1$, A. S. Rastorguev$^{2,3}$, and M. V. Zabolotskikh$^{2}$\\
     $^1$Central (Pulkovo) Astronomical Observatory of RAS, 65/1
     Pulkovskoye Chaussee, Saint Petersburg, 196140, Russia\\
     $^2$Sternberg Astronomical Institute, Lomonosov Moscow State University, 13 Universitetskii prospest, Moscow, 119992, Russia\\
     $^3$Faculty of Physics, Lomonosov Moscow State University, 1 bldg.2, Leninskie Gory, Moscow, 119991, Russia}

\begin{document}
 \date{Accepted 2021 January 07. Received 2021 January 07; in original form 2020 May 11}
 \pagerange{\pageref{firstpage}--\pageref{lastpage}} \pubyear{2021}
 \maketitle
 \label{firstpage}

\begin{abstract}
To construct the rotation curve of the Galaxy, classical Cepheids with proper motions,  parallaxes and line-of-sight velocities from the Gaia DR2 Catalog are used in large part. The working sample formed from literature data contains about 800 Cepheids with estimates of their age. We determined that the linear rotation velocity of the Galaxy at a solar distance is $V_0=240\pm3$~km s$^{-1}$. In this case, the distance from the Sun to the axis of rotation of the Galaxy is found to be $R_0=8.27\pm0.10$~kpc. A spectral analysis of radial and residual tangential velocities of Cepheids younger than 120 Myr showed close estimates of the parameters of the spiral density wave obtained from data both at present time and in the past. So, the value of the wavelength $\lambda_{R,\theta}$  is in the range of [2.4--3.0] kpc, the pitch angle $i_{R,\theta}$ is in the range of [$-13^\circ$,$-10^\circ$] for a four-arm pattern model, the amplitudes of the radial and tangential perturbations are $f_R\sim12$~km s$^{-1}$ and $f_\theta\sim9$~km s$^{-1}$, respectively. Velocities of Cepheids older than 120 Myr are currently giving a wavelength $\lambda_{R,\theta}\sim5$~kpc. This value differs significantly from one that we obtained from the samples of young Cepheids.  An analysis of positions and velocities of old Cepheids, calculated by integrating their orbits backward in time, made it possible to determine significantly more reliable values of the parameters of the spiral density wave: wavelength $\lambda_{R,\theta}=2.7$~kpc, amplitudes of radial and tangential perturbations are $f_R=7.9$~km s$^{-1}$ and $f_\theta=5$~km s$^{-1}$, respectively.
  \end{abstract}


 \section{INTRODUCTION}
Cepheids are of great interest because they implement an independent scale of astronomical distances. For these variable stars it is possible thanks to the period--luminosity \citep{Leavitt1908,Leavitt1912} (PLR) and period--Wesenheit \citep{Madore1982, Caputo2000} relations (PWR).  Currently, these relations are well calibrated using high-precision trigonometric parallaxes of stars \citep{Ripepi19}. The use of this relations allows us to estimate the distances to Cepheids with random errors smaller than 10\%
\citep{Berd00,Sandage06,Skowron19}. Note that \cite{Lazovik20}
derived PLR by new method, using multiphase temperature
measurements, which made it possible to calculate the most
accurate individual color excesses of Cepheids used. The method is
based on the Baade-Becker-Wesselink approach, and practically does
not use trigonometric parallaxes.

Classical Cepheids are young (under $\sim$400 million years old)
supergiant stars  with periods of radial pulsations from $\sim$1
to $\sim$100 days. They are attributed to the flat component of
the stellar population of the Galaxy, therefore they are used to
study the structural and kinematic features of the galactic disk.

In the works of various authors \citep{Joy1939,Pont1997,Metzger1998},
the rotation  parameters of the Galaxy were determined using only the
distances and line-of-sight velocities of Cepheids.

To determine the  Galactic rotation parameters, the combination of distances,  line-of-sight velocities, and proper motions of Cepheids  was used, for example, in \cite{Frink95}. In this case proper motions from the PPM \citep{Roeser88} catalog, which are not very accurate, were taken.

On the basis of classical Cepheids with  proper motions from the Hipparcos Catalog~\citep{1997}, there were refined the galactic rotation parameters \citep{Feast97, Melnik2015}, the parameters of the spiral structure \citep{Melnik1999, BobylevBaj12, Dambis15} and the parameters of the galactic disk   flexure\citep{Bobylev13a,Bob13b}. Based on Cepheids of II~type (the old, low mass counterpart to classical Cepheids), there were determined the parameters of the central bulge and the distance to the galactic center  \citep{Majaess09}.

The measurements obtained by the space experiment Gaia ~ \ citep {Prusti16} are unprecedented in accuracy and volume for the study of the Galaxy. Currently the second version of the Catalog, Gaia DR2, has been published \citep{Brown18}. The average errors of
trigonometric parallaxes of bright stars ($G<15^m$) in this
Catalog lie in the range of 0.02--0.04~milliarcseconds (mas), and
for faint stars ($G=20^m$) they are of the order 0.7~mas. Similarly, the
proper motion errors vary from 0.05~mas year$^{-1}$ for bright
($ G<15^m $) to 1.2~mas year$^{-1}$ for faint ($G=20^m$) stars.
Line-of-sight velocities of more than 7 million stars are
measured. For stars of spectral classes F-G-K, the average error
of the line-of-sight velocities is about 1 km s$^{-1}$.

The problem of establishing the zero point of parallaxes in the Gaia DR2 Catalog is known. Already \cite{Lindegren18} has indicated the presence of a possible systematic parallax
zero-point offset of $\Delta\pi=-0.029$~mas in Gaia DR2 relative
to the inertial reference frame. Currently, there are several
reliable independent estimates of this offset. So, from a
comparison of eclipsing binary stars, \cite{Stassun18} found
$\Delta\pi=-0.082\pm0.033$~mas. This value is confirmed by other
authors, in particular, in the analysis of Cepheids
$\Delta\pi=-0.046\pm0.013$~mas \citep{Riess18a},
$\Delta\pi=-0.049\pm0.018$~mas \citep{Groenewegen18},
$\Delta\pi=-0.071\pm0.038$~mas \citep{Skowron19} and
asteroseismology $\Delta\pi=-0.053\pm0.009$~mas \citep{Zinn19}.
PMZ and MZ relations were derived by hierarchical Bayesian
approach for approximately 400 RR~Lyrae stars with optical and NIR
photometry and Gaia DR2 data \citep{Muraveva18} to give $\Delta\pi
\approx -(0.54-0.62)$~mas. Note that the work \cite{Riess18a} used
50 long-period Cepheids with high-precision photometry performed
with the Hubble Space Telescope. In the work \cite{Groenewegen18},
a sample of 452 Classical Cepheids was used and 251 Classical
Cepheids was used in the work \cite{Skowron19}.

\cite{Skowron19} built a three-dimensional map of the distribution
of  2431 Cepheids in the Galaxy. For this, the classical Cepheids
of the main program OGLE (Optical Gravitational Lensing
Experiment, \cite{Udalski97}), were supplemented by Cepheids from
GCVS (General Catalog of Variable Stars, \cite{Samus17}), ASAS
(All Sky Automated Survey, \cite{Pojmanski02}), and  by about 200
Cepheids from the Gaia DR2 Catalog and a number of other sources.
Using such biggest sample, these authors  specified the parameters
of the density distribution of galactic Cepheids using the
exponential law, and the parameters of the warped galactic disk.
We note their Fig.~3 and Fig.~4, from which it is seen how young
Cepheids trace a galactic spiral pattern.

In the work of \cite{Mroz19} performed on 773 classical Cepheids
with proper motions and line-of-sight velocities from the Gaia DR2
Catalog, the galactic rotation parameters were determined with the
highest accuracy. In particular, the galactic rotation velocity at
a solar distance was found to be $V_0=233.6\pm2.8$~km s$^{-1}$,
and its first derivative $V'_0=-1.34\pm0.21$~km s$^{-1}$
kpc$^{-1}$. As shown by \cite{BobylevBaj12}, some of the spiral density wave
parameters depend on the Cepheids age. Such a parameter is, for example, the phase
of the Sun in a spiral wave. Therefore, it is interesting to
determine such parameters using Cepheids of different ages from the
latest data.

When analyzing maser sources, \cite{Rastorguev17} obtained the rotation curve parameters in
combination with the Str$\ddot{o}$mberg asymmetry parameters, which allowed us to
estimate the exponential scale of the galactic disk under the assumption of marginal stability of the intermediate-age disk. It is interesting to compare the results of this approach in application to the kinematic and position data of a large sample of Cepheids.

In the work of \cite{Gnacinski19} it has been shown that the rotation velocities of classical Cepheids, obtained in three ways: 1)only from line-of-sight velocities, 2)only from proper motion (Gaia DR2) and 3)from full three-dimensional velocity vector, are located between the flat and Keplerian rotation curves. Using a large sample of Cepheids, \cite{Ablimit20} estimated the rotation velocity of the Galaxy at the Sun position and found the virial mass of the Galaxy and local dark matter density.

Using large number of stars from Gaia DR2, \cite{Kawata18} have generated the maps of the rotation velocity, $V_{circ},$ and vertical velocity, $V_z,$ distributions as a function of the Galactocentric radius, $R$. In the $R-V_z$ distribution they found the peak of the $V_z$ distribution shows wave-like features.

Cepheids, as unique ``standard candles'', play an important role in the construction of a universal distance scale due to the presence of the period--luminosity relations. These high-luminosity stars can be detected and studied with large ground-based and space telescopes in disk galaxies up to the distances of 20--30 Mpc. Cepheid distances up to several tens of galaxies, where supernova explosions of type Ia were recorded, have long served as the basis for calibrating the luminosities of these supernovae at their maximum brightness. As a consequence, Hubble diagrams for type Ia cosmological supernovae led to the
discovery of the accelerated expansion of the Universe (``dark energy'', see
 \cite{Riess1998,Perlmutter1997,Riess2004,Perlmutter1999,Schmidt1998} and others).

The combined use of Hubble diagrams for type Ia cosmological supernovae, data on the CMB anisotropy measured by WMAP and Planck space missions, and the results of study of large-scale distribution of galaxies (BAO -- Barionic Acoustic Oscillations)
made it possible to set the restrictions on the values of the global cosmological parameters: the contribution of the baryonic and non-baryonic matter and that of the dark energy to the total mass-energy density, as well as the curvature parameter and the
equation of state (see the most important recent papers by
\cite{Betoule14, Abbott19, Scolnic18}).

In the last 5--6 years, strong evidence appeared in favor of a significant discrepancy between the values of the Hubble constant $H_0$, determined from the CMB anisotropy and from the redshifts of galaxies and brightest optical ``standard candles'' -- Type Ia supernovae, whose luminosities are based on the Cepheid distances of galaxies. Cepheid methods lead to systematically higher $H_0$ values by about 6--7 km s$^{-1}$ Mpc$^{-1}$ at a significance
level of more than 4$\sigma$. This problem is now well known as ``Hubble tension'' (see, for example, \cite{Riess2018b, Verde2019,Riess2020} and references therein).

 \begin{figure*}
 {\begin{center}
 \includegraphics[width=155mm]{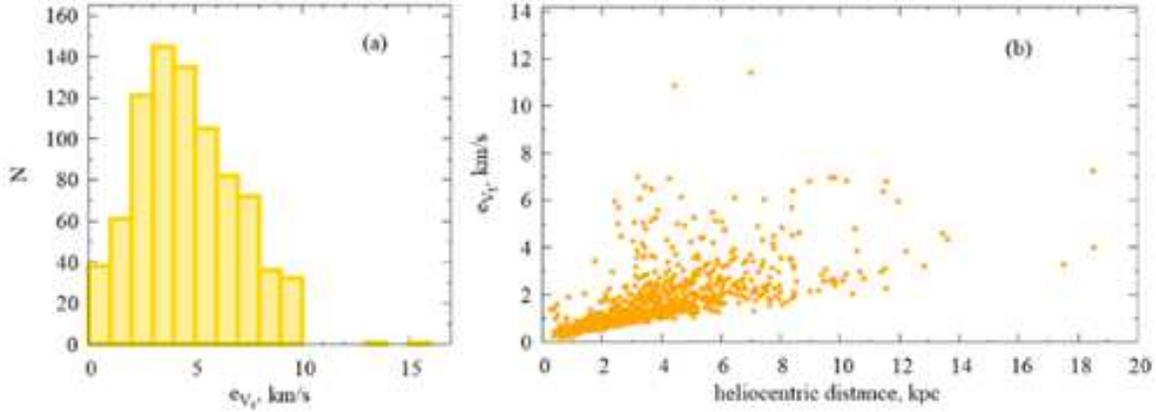}
 \caption{
A histogram of random errors of line-of-sight velocities of Cepheids~(a); errors of the tangential velocities of these stars versus the heliocentric distance~(b). }
 \label{f-er}
 \end{center} }
 \end{figure*}

Note that the group of CMB-based estimates is often referred to as ``global'', referring to the early Universe, while the Cepheid-based group is called as ``local'', referring to the
recent-epoche Universe. The reasons for the differences and the ways of solving the ``tension'' problem are being actively discussed (for example, we found about 750 articles with the mention of the term ``Hubble tension'' in the abstracts of the
papers published during last 5 years). In most works, the problem
of Hubble tension is analyzed from the theoretical point of view
concerned the properties of dark energy and the refinement of the
existing models of the Universe.

However, it is possible that the reason for the Hubble tension can be partly explained by the existence of some systematical errors in the Cepheid distance scales used, though the random relative errors of Cepheid distances are within 10\%. The main reason for
possible systematics is commonly attributed to the differences in
the metallicity of galaxies hosting Cepheids. \cite{Sandage06}
discussed this issue in details, but this question is still under
numerous debates. The second reason -- the systematic errors of
different kinds that arise during calibration of PLR by
trigonometric parallaxes; some are due to nonlinear conversion of
parallaxes to distances, and some -- to parallax zero-point offset
inherent in the Gaia DR2/EDR3 catalogs discovered in a large
number of studies (see discussion above). The PLRs derived by the
Baade-Wesselink technique can also be distorted by systematic
errors. Additional source of possible errors in the Hubble
constant determinations based on the redshifts of galaxies and
supernovae stars is an underestimation of the influence of the
velocity dispersion of galaxies in galaxy clusters (see, for
example, \cite{Sedgwick21}). All these issues affecting the
accuracy of the distance scale, including systematic effects, are
discussed in the literature, but for a detailed analysis of the
kinematics of the Galactic Cepheids presented in this paper, they
do not constitute a serious problem.

The aim of this work is to estimate the rotation parameters of the
Galaxy, as well as spiral density wave parameters using a large
sample of classical Cepheids of different age with proper motions
and line-of-sight velocities taken from the Gaia DR2 Catalog.

\section{Data}\label{data}
In this work, we use the data on Classical Cepheids from the works of \cite{Mroz19} and \cite{Skowron19}.

The Catalog of \cite{Skowron19} contains distance, age, pulsation period, and the mid-infrared (mid-IR) data from Spitzer \citep{Benjamin03,Churchwell09} and WISE space telescopes \citep{Wright10,Mainzer11} for 2431 Cepheids. The distances to these stars, $r$, were calculated by \cite{Skowron19} on the base of MIR PLR of \cite{Wang18} and
mid-infrared light curves, where the influence of the interstellar
absorption is much smaller than in optics.

There is a debate concerning period-age relations \citep{Turner12}.
There are known several calibrations proposed to estimate Cepheid's mean age. For example,
the theoretical calibration performed by \cite{Bono05} and the calibration by \cite{Efremov03}, obtained by analysis of Cepheids in the Large Magellanic Cloud. To estimate the age of Cepheids \cite{Skowron19} used the calibration from the paper of \cite{Anderson2016}.

We note that such a fundamental property of stars as the rotation \citep{Anderson2016} is of great importance for the study of the classical Cepheid variable stars. The models by \cite{Anderson2016} include rotation, while some other stellar models \citep{Bono05}
do not, what makes the Cepheids 1.5 to 2 times younger. It should also be noted that the ages derived by \cite{Skowron19} were extrapolated from \cite{Anderson2016} for the metal-rich Cepheids.

The catalog of \cite{Mroz19} contains data on 832 Classical
Cepheids. The proper motions and line-of-sight velocities of stars
included to the catalog, with appropriate errors, are taken from
the Gaia DR2 Catalog. We supplemented the Catalog of \cite{Mroz19}
with estimates of the age of Cepheids from the work of
\cite{Skowron19}.

 \begin{figure*}  \begin{center}
 \includegraphics[width=150mm]{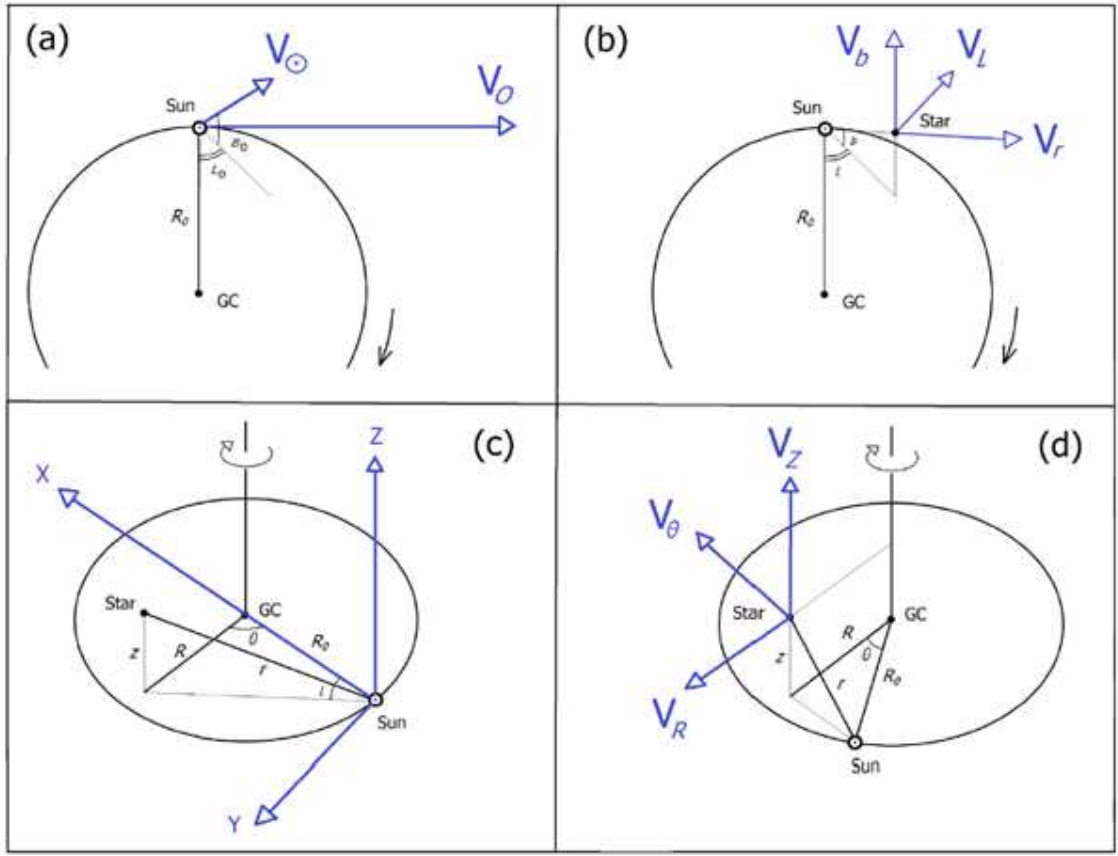}
 \caption{
The peculiar velocity of the Sun $V_\odot$ and circular rotational
velocity of the Sun $V_0$ around Galactic Center (GC) at distance
$R_0$ (a), the velocity components $V_r, V_l$ and $V_b$ (b),
coordinate system $x,y$ and $z$ (c), cylindrical coordinate system
$V_R, V_\theta$ and $V_z$(d), thin black arrows indicate the
direction of rotation of the Galaxy.}
 \label{f-99099}
 \end{center}
 \end{figure*}

Heliocentric distances to Cepheids were taken from 76--81 columns of the Catalog of \cite{Mroz19}. These, in turn, are taken from the work of \cite{Skowron19}, where
they were calculated using MIR PLR of \cite{Wang18} and
mid-infrared light curves, which virtually removes the effects of
interstellar extinction.

Apparent stellar magnitudes of Cepheids observed in OGLE program, lie in the range from $I=11^m$ to $I=18^m$ \citep{Skowron19}. Therefore, in \cite{Mroz19} and \cite{Skowron19} catalogs there is a deficiency of bright and well-studied Cepheids from earlier
observations.

According to~\cite{Skowron19} the errors in distances to the Cepheids are $\sim$5\%.
Problem is that there may be systematics at play toward certain directions, and at given short and long Galactocentric radii where metallicity and $R$ (ratio of the total-to-selective extinction) differences may be at play, etc.

Random line-of-sight velocity errors $e_{V_r}$ usually do not exceed 10~km s$^{-1}$, in average they are 5~km s$^{-1}$. Typical proper motion error of about 0.1~mas year$^{-1}$ will give tangential velocity error $e_{V_t}=5$~km s$^{-1}$ ($0.1\cdot
4.741\cdot r$) only for heliocentric distances greater than
10~kpc. Thus, in our sample, random line-of-sight velocity errors
introduce the main contribution to random errors of spatial
velocities.

The errors of distances and line-of-sight velocities have been taken from the Catalog of \cite{Mroz19}. Fig.~\ref{f-er} represents a histogram of random line-of-sight velocity errors $e_{V_r}$ of Cepheids (left-hand panel), and errors of
tangential velocities $e_{V_t}$ of these stars versus the
heliocentric distance (right-hand panel). It can be seen from the
figure that our approximate estimate is in good agreement with the
actual distribution of Cepheid's random velocity errors.

\section{Method}\label{method}
\subsection{The Galaxy rotation curve parameters}
We use a rectangular coordinate system centered on the Sun. The $x$~axis is directed
towards the Galactic center, the direction of the $y$~axis coincides with the direction of rotation of the Galaxy, and the $z$~axis is directed towards the north pole of the Galaxy.
Then the rectangular coordinates are calculated as follows:
$x=r\cos l\cos b,$ $y=r\sin l\cos b$ and $z=r\sin b.$
This coordinate system is shown in Fig.~\ref{f-99099}c.

Astrometric observations give three components of a star velocity: the line-of-sight velocity $V_r$ and two projections of the tangential velocity: $V_l=4.74r \mu_l\cos b$ and $V_b=4.74r\mu_b$ directed along the galactic longitude $l$ and latitude $b$ respectively. All the velocity components are measured in km s$^{-1}$. Proper motion components $\mu_l\cos b$ and $\mu_b$ are determined in mas year$^{-1}$. The coefficient 4.741 is equal to the ratio of the number of kilometers in an astronomical unit to the number
of seconds in a tropical year, and $r$ is the star's heliocentric
distance in kpc.  The velocities $V_r,V_l,$ and $V_b,$ are shown in Fig.~\ref{f-99099}b.

For stars with known line-of-sight velocities, proper motions and
distances, the spatial velocities $U,V,W$ are calculated as follows:
 \begin{equation}
 \begin{array}{lll}
 U=V_r\cos l\cos b-V_l\sin l-V_b\cos l\sin b,\\
 V=V_r\sin l\cos b+V_l\cos l-V_b\sin l\sin b,\\
 W=V_r\sin b                +V_b\cos b.
 \label{UVW}
 \end{array}
 \end{equation}
Herewith the velocity $U$ is directed from the Sun to the
galactic center, $V$ is in the direction of galactic rotation
and $W$ is directed to the north galactic pole.

In further studies of the galactic spiral density wave, we also use the following two very important velocities: the radial velocity $V_R$, directed from center of the galaxy to the star, and the tangential velocity $V_{circ}$, orthogonal to $V_R$ and directed towards the rotation of the Galaxy, which are calculated using the following formulas:
 \begin{equation}
 \begin{array}{lll}
  V_{circ}= U\sin \theta+(V_0+V)\cos \theta, \\
       V_R=-U\cos \theta+(V_0+V)\sin \theta,
 \label{VRVT}
 \end{array}
 \end{equation}
where $R_0$~is the galactocentric distance of the Sun, $V_0$~is
the linear circular rotation velocity around the galactic center
in the solar neighbourhood, $R$~is the distance from the
star to the axis of galactic rotation:
  \begin{equation}
 R^2=r^2\cos^2 b-2R_0 r\cos b\cos l+R^2_0,
  \end{equation}
and the position angle $\theta$ meets the relation $\tan\theta=y/(R_0-x)$.
Components $V_R,$ $V_\theta$ (in our case $V_\theta\equiv V_{circ}$) and $V_z,$ are shown
later in Fig.~\ref{f-99099}d).

We determine the parameters of the galactic rotation curve by solving equations based on Bottlinger's formulas, in which the angular velocity $\Omega$ is expanded into a Taylor series in powers of $(R-R_0)$ to the terms of the $i$th order of smallness of $r/R_0$:
\begin{equation}
 \begin{array}{lll}
  V_r=-U_\odot\cos b\cos l-V_\odot\cos b\sin l-W_\odot\sin b\\
      +R_0\sin l\cos b\left[\sum\limits_{i=1}^N(R-R_0)^i
      {\displaystyle\Omega_0^{(i)}\over \displaystyle i!}\right],
 \label{EQ-1}
 \end{array}
 \end{equation}
 \begin{equation}
 \begin{array}{lll}
   V_l=U_\odot\sin l-V_\odot\cos l-r\Omega_0\cos b\\
 +(R_0\cos l-r\cos b)\left[\sum\limits_{i=1}^N(R-R_0)^i
 {\displaystyle\Omega_0^{(i)}\over \displaystyle i!}\right],
 \label{EQ-2}
 \end{array}
 \end{equation}
 \begin{equation}
 \begin{array}{lll}
  V_b=U_\odot\cos l\sin b+V_\odot\sin l\sin b-W_\odot\cos b\\
  -R_0\sin l\sin b\left[\sum\limits_{i=1}^N(R-R_0)^i
  {\displaystyle\Omega_0^{(i)}\over \displaystyle i!}\right].
 \label{EQ-3}
 \end{array}
 \end{equation}
The values $U_\odot, V_\odot$ and $W_\odot$ is a group velocities which contain the peculiar motion of the Sun (see Fig.~\ref{f-99099}a) and the contribution from the effect called
``asymmetric drift'', which is considered to be small in the case
of Cepheids and other young populations. The value $\Omega_0$ is
the angular velocity of the Galaxy at a solar distance $R_0$,
$\Omega_0^{(i)}$~is the $i$-th derivative of the angular velocity
with respect to $R$, the linear rotation velocity at a solar
distance equals to $V_0=R_0\Omega_0$. In the coordinate system
$x,y,z$ shown in Fig.~\ref{f-99099}c with positive rotation around
the $z$ axis, there will be a rotation from the $x$ axis to $y$.
In this case, the sign of the angular velocity $\Omega$ will be
negative, which is not always convenient. We prefer to have a
positive $\Omega$. Therefore, the equations
(\ref{EQ-1})--(\ref{EQ-3}) are written appropriately (such a
coordinate system is shown later). 

\subsection{Residual Velocity Formation}\label{residual}
The residual velocities are calculated taking into account the peculiar motion of the Sun,
$U_\odot, V_\odot$ and $W_\odot$ (see Fig.~\ref{f-99099}a), as well as the influence of the differential rotation of the Galaxy in the following way:
\begin{equation}
 \begin{array}{lll}
 V_r=V^*_r-[-U_\odot\cos b\cos l-V_\odot\cos b\sin l-W_\odot\sin b\\
 +R_0(R-R_0)\sin l\cos b\Omega^\prime_0\\
 +0.5R_0(R-R_0)^2\sin l\cos b\Omega^{\prime\prime}_0+\ldots],
 \label{EQU-1}
 \end{array}
 \end{equation}
 \begin{equation}
 \begin{array}{lll}
 V_l=V^*_l-[U_\odot\sin l-V_\odot\cos l-r\Omega_0\cos b\\
 +(R-R_0)(R_0\cos l-r\cos b)\Omega^\prime_0\\
 +0.5(R-R_0)^2(R_0\cos l-r\cos b)\Omega^{\prime\prime}_0+\ldots],
 \label{EQU-2}
 \end{array}
 \end{equation}
  \begin{equation}
 \begin{array}{lll}
 V_b=V^*_b-[U_\odot\cos l\sin b + V_\odot\sin l \sin b -W_\odot\cos b\\
 -R_0(R-R_0)\sin l\sin b\Omega^\prime_0\\
 -0.5R_0(R-R_0)^2\sin l\sin b\Omega^{\prime\prime}_0-\ldots],
 \label{EQU-3}
 \end{array}
 \end{equation}
where $V^*_r,V^*_l,V^*_b$~standing on the right-hand sides of the equations are the initial velocities, and on the left-hand sides there are the corrected velocities $V_r,V_l,V_b$~which can be used for calculation of the residual velocities $U,V,W$ by the formulas (\ref{UVW}).

 \begin{figure*} {\begin{center}
 \includegraphics[width=140mm]{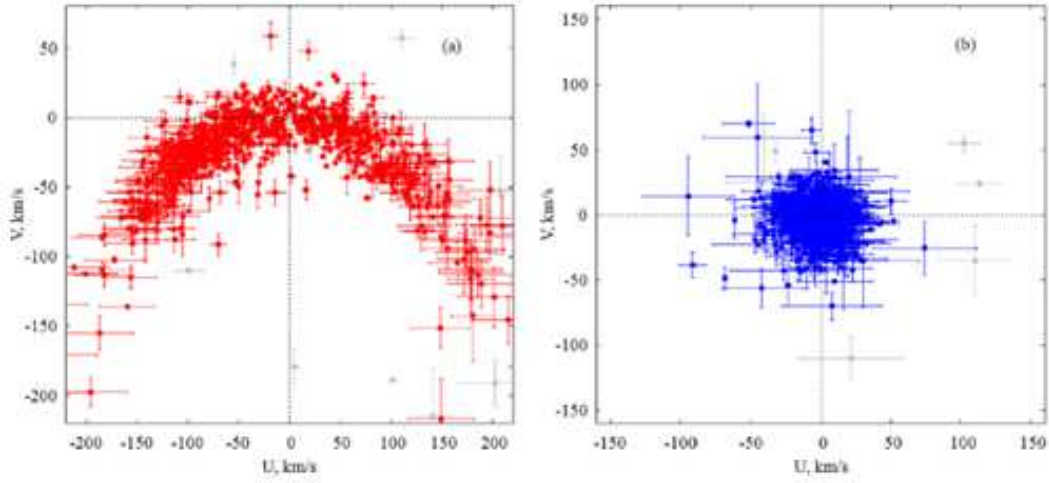}
 \caption{Cepheid's $U,V$ velocities: not corrected (a) and corrected (b)
for the differential galactic rotation; gray symbols indicate
stars that were discarded according to the restrictions
(\ref{cut}).}
 \label{f-UV}
 \end{center} }
 \end{figure*}

The original spatial velocities of stars usually contain a small percentage of values that
differ significantly from the average. Such velocities should be discarded using some criterion. Fig.~\ref{f-UV} shows the Cepheids $U,V$ velocities calculated
using the relations~\ref{UVW}. All Cepheids (820 stars) were used
to build this picture without any preliminary rejection. We see
that before analyzing such velocities in order to detect rebounds,
they must be corrected for the differential rotation of the
Galaxy.

\begin{figure*}
 \begin{center}
  \includegraphics[width=120mm]{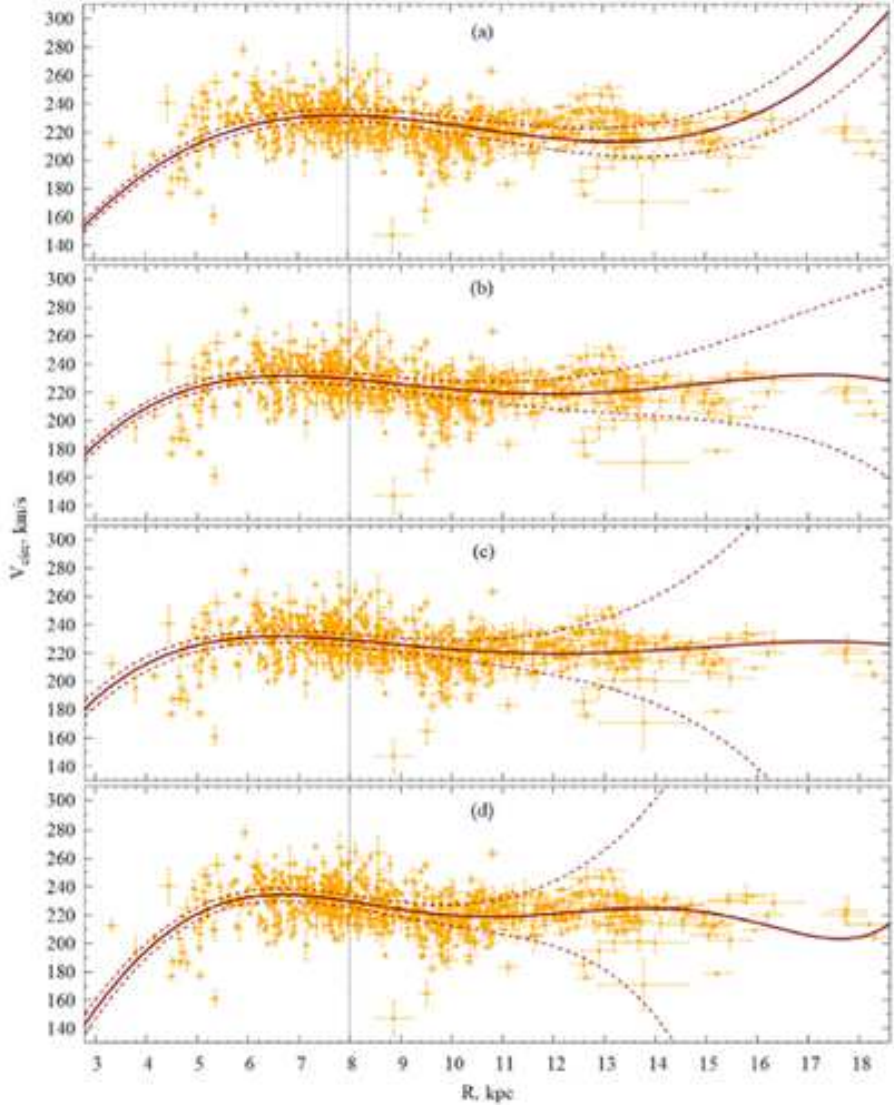}
 \caption{
The rotation velocities of Cepheids $V_{circ}$ versus the distance
$R$, the galaxy rotation curve found from these stars with two
derivatives of the angular velocity of rotation (a), with three
derivatives (b), with four derivatives (c) and with five
derivatives (g); for each curve, the confidence interval limits
corresponding to the error level of $1\sigma$ are marked with
dashed lines; all curves are calculated for the accepted value of
$R_0=8$~kpc. The vertical line marks the position of the Sun.
 }
 \label{f1-rot-3456}
 \end{center}
\end{figure*}

Note that the application of the rotation curve obtained on the
basis of the equations~(\ref{EQ-1})--(\ref{EQ-3}) has a
restriction on $R$ (Fig.~\ref{f1-rot-3456}). Therefore, for
purposes such as detecting bounces when analyzing the residual
tangential velocities $|\Delta V_{circ}|$, we can simply use the
flat rotation curve, $V_{circ}=const$. The velocities $W$ and
$V_R$ do not depend on the rotation curve of the Galaxy (see
(\ref{VRVT})). As a result, to remove bounces from the sample, we
apply the following restrictions:
 \begin{equation}
 \begin{array}{rcl}
             |V_R|<90~\hbox{km s$^{-1}$},\\
 |\Delta V_{circ}|<90~\hbox{km s$^{-1}$},\\
               |W|<60~\hbox{km s$^{-1}$}.
  \label{cut}
 \end{array}
 \end{equation}
Note that the criteria~(\ref{cut}) are only used for a preliminary
cleaning of the sample, while another rejection criterion
(velocity residuals in excess of 3 sigma) will be applied in the
analysis.

It is known that the movement of gas and young stars is influenced
by the Central galactic bar~\citep{Chemin19}. Some authors refer
to the bar as a structure with a half length of 2.5~kpc with a
position angle of 15--30 degrees with respect to the Sun-Galactic
Centre direction~\citep{Babusiaux05,Lopez05}, other researchers
suggest that there is a long massive bar with a half length of
4--5~kpc and a position angle of around 45 degrees
\citep{Hammersley1994,Wegg15}. Moreover, according to some authors \citep{Alard01,Nishiyama05}, there is also a very small inner bar embedded to very central
bar/buldge structure with a different orientation as compared to
the other two bars.

\begin{figure*}
  \begin{center}
\includegraphics[width=140mm]{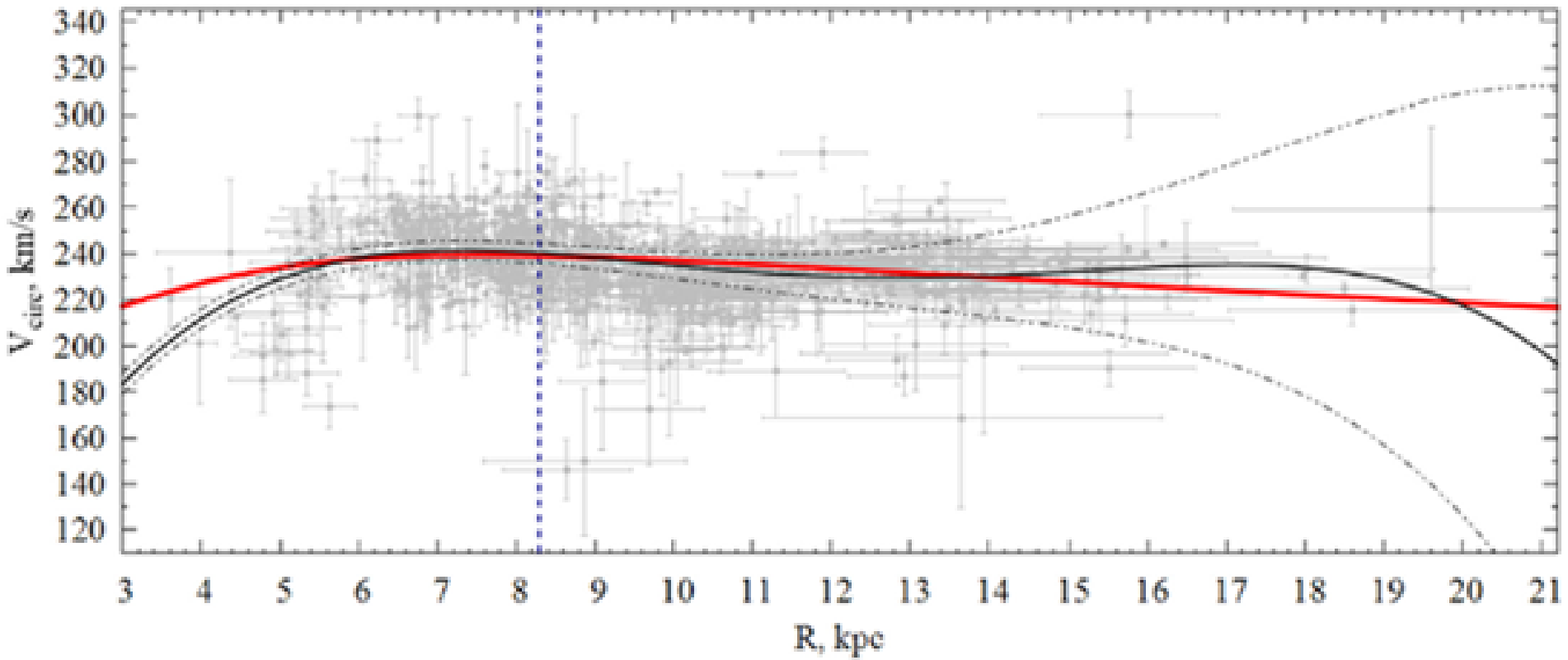}
 \caption{
The rotation velocities of Cepheids $V_{circ}$ versus the distance
$R$; the thin solid line shows the Galaxy rotation curve found
from these stars (solution~(\ref{solution-all})), the wide red
line shows the rotation curve corresponding to the potential
model~III~\citep{Bajkova16}. The confidence interval limits
corresponding to the error level of $1\sigma$ are marked with
dashed lines. The vertical line marks the position of the Sun. }
\label{f1-2rot}
 \end{center}
 \end{figure*}

The strongest influence (deviations from the flat rotation curve
more than $\sim$50~km s$^{-1}$, an increase in velocity
dispersions) on the velocity of objects due to the bar is observed
in the region of 1.5--2~kpc
\citep{Clemens1985,Bhattacharjee14,Bajkova16}, a weak
gravitational influence can be traced up to 5~kpc
\citep{Chemin19}. To exclude the influence of the bar, it is
advisable to apply the restriction $R>2.5$~kpc as an initial
condition. As can be seen from Fig.~\ref{f1-rot-3456} and
Fig.~\ref{f1-2rot}, practically almost all the Cepheids we use for
the kinematical analysis lie at the distances $R>4$~kpc.

In the Catalog of \cite{Mroz19}, some Cepheids are located very far
from the Galaxy center, at $r>25$~kpc, which are also better to
exclude from consideration since at large distances proper motion
error of about 0.1~mas year$^{-1}$ will lead to tangential
velocity error $e_{V_t}>10$~km s$^{-1}$ (see Fig.~\ref{f-er}b). As
a result of the above restrictions, no more than 20 stars are
discarded. The total sample contains 800 Cepheids.


\subsection{The Spectral Analysis}
According to the linear theory \citep {Lin1964}, the influence of the spiral density wave on the radial $V_R$ and residual tangential velocities $\Delta V_{circ}$ has the character of a periodic functions and is described by the following relations:
 \begin{equation}
 \begin{array}{rcl}
       V_R =-f_R \cos \chi,\\
 \Delta V_{circ}= f_\theta \sin\chi,
 \label{DelVRot}
 \end{array}
 \end{equation}
where  $f_R$ and $f_{\theta}$~are positive definite amplitudes of the perturbations of the radial and residual tangential velocities, respectively;
  \begin{equation}
 \chi=m[\cot(i)\ln(R/R_0)-\theta]+\chi_\odot
 \end{equation}
is the phase of the spiral wave, where $m$~is the number of spiral arms, $i$~is a pitch angle of the spiral pattern, $\chi_\odot$~is a radial phase of the Sun in a spiral wave.
As an analysis of modern high-precision data showed, the periodicities
associated with a spiral density wave also appear in vertical
velocities $W$ \citep{BobylevBaj15,Rastorguev17}.

To identify periodicities in the velocities $V_R$ and $\Delta V_{circ}$, we use a modified spectral (periodogram) analysis \citep{Bajkova12}. Wavelength $\lambda$ (distance between adjacent pieces of spiral arms, counted along the radial direction) is calculated as
follows:
\begin{equation}
 \frac{2\pi R_0}{\lambda}=m\cot(i).
 \label{a-04}
\end{equation}

Let there be  a series of measured velocities $V_{R_n}$ (these can be $V_R$ or $\Delta V_{circ}$ velocities), $n=1,\dots,N$, where $N$~is a number of objects. The task of spectral analysis is to extract the periodicity from a data series in accordance with the accepted model describing a spiral density wave with parameters $f,\lambda$~(or $i$) and $\chi_\odot$.

As it was shown by \cite{Bajkova12}, taking into account the logarithmic nature of the spiral density wave, as well as the positional angles $\theta_n$ of objects, our spectral analysis of the series of velocities can be reduced to calculating of the power spectrum of the standard Fourier transform :
\begin{equation}
 \bar{V}_{\lambda_k} = \frac{1} {N}\sum_{n=1}^{N} V^{'}_n(R^{'}_n)
 \exp\biggl(-j\frac {2\pi R^{'}_n}{\lambda_k}\biggr),
 \label{29}
\end{equation}
where $\bar{V}_{\lambda_k}$~is the $k$-th harmonic of the Fourier
transform with the wavelength $\lambda_k=D/k$, $D$~is the period
of the analyzed series;
 \begin{equation}
 \begin{array}{lll}
 R^{'}_{n}=R_0\ln(R_n/R_0),\\
 V^{'}_n(R^{'}_n)=V_n(R^{'}_n)\times\exp(jm\theta_n).
 \label{21}
 \end{array}
\end{equation}
The peak value of the power spectrum $S_{peak}$ corresponds to the
desired wavelength $\lambda$. The pitch angle of the spiral
density wave can be found from~(\ref{a-04}). We find
the amplitude and phase of the perturbations as a result of
fitting the harmonics with the found wavelength to the measured
data. To estimate the amplitude of disturbances, we use the
relation:
 \begin{equation}
 f_R(f_\theta)=2\times\sqrt{S_{peak}}.
 \label{Speak}
 \end{equation}

 \begin{figure*}{\begin{center}
 \includegraphics[width=150mm]{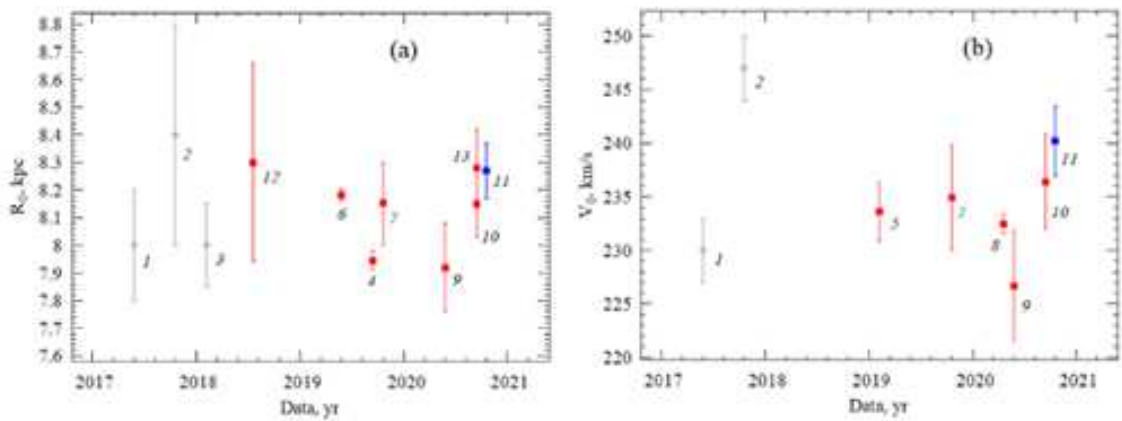}
 \caption{
The results of determining the distance $R_0$ ~(a) and velocity
$V_0$ ~(b) by various authors depending on the date of
publication, gray color indicates the results obtained as an
average, red --- individual determinations, blue --- the result of
this work, see also the text.}
 \label{f-R0V0}
 \end{center}
  }
 \end{figure*}

\subsection{The choice of the $R_0$ value}
Currently, a number of works devoted to determining the average value of the Sun galactocentric distance have been performed using individual definitions of this quantity, obtained in the last decade by independent methods.

We note several important results derived as an average over a
large number of independent estimates of $R_0$. For instance,
 $R_0=8.0\pm0.2$~kpc \citep{Vallee17a},
 $R_0=8.3\pm0.4$~kpc \citep{Grijs17} or
 $R_0=8.0\pm0.15$~kpc \citep{Camarillo18}.

We also note some of the first-class individual definitions of this quantity made recently. In the work of \cite{Abuter19}, from the analysis of a 16-year-long series of observations of the motion of the S2 star around a super-massive black hole in the center of the Galaxy, the value $R_0=8.178\pm0.022$~kpc was found. In the work of \cite{Do19}, based on an independent analysis of the orbit of star S2, the value $R_0=7.946\pm0.032$~kpc was found.
Using  data on galactic masers obtained with the Japanese program VERA (VLBI Exploration of Radio Astrometry)  received an estimate of $R_0=7.9\pm0.3$~kpc \citep{Hirota2020}. Estimates obtained from the analysis of variable stars are also of interest. From the analysis of VVV-based (VISTA Variables in the Via Lactea) near-infrared RR~Lyrae data \cite{Majaess18} obtained $R_0=8.30\pm0.36$~kpc. From the analysis of OGLE-based RR Lyrae data \cite{Griv20} obtained $R_0=8.28\pm0.14$~kpc.

Based on the above results, in the present work we assume the value $R_0=8.0\pm0.15$~kpc in cases where $R_0$ is not a definable parameter.

 \begin{figure*}  \begin{center}
 \includegraphics[width=140mm]{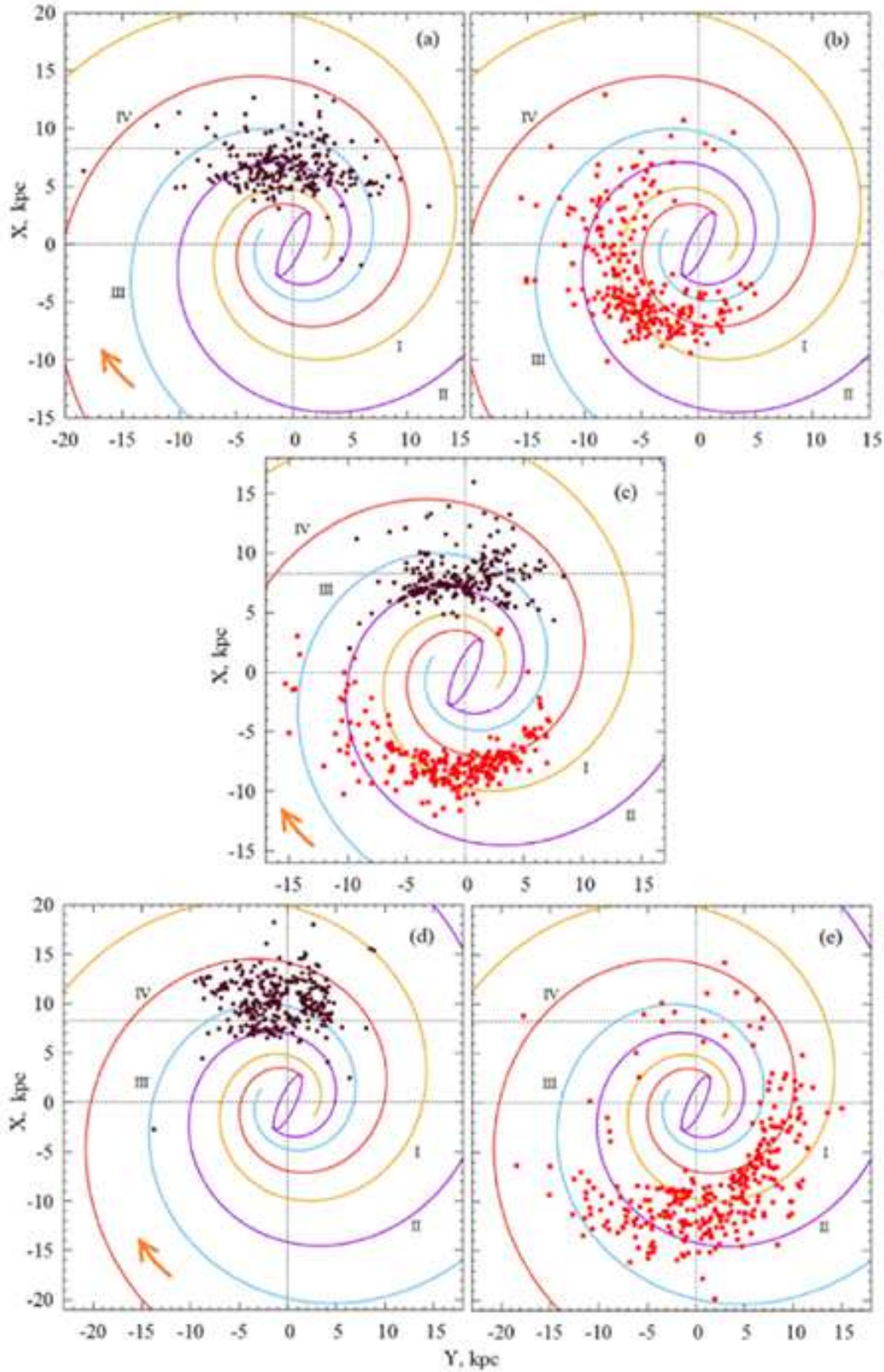}
\caption{The distribution on the galactic plane $X,Y$ of the
youngest Cepheids at the present time (a) and in the past
according to the age of each star ~(b), middle-aged Cepheids (~c),
samples of old Cepheids at the present time ~(d) and in the past
~(e), a four-arm spiral pattern with a pitch angle of
$-13^\circ$~\citep{BobBaj14} is given, the spiral arms are
numbered in Roman numerals, the orange arrow shows the direction
of the Galaxy rotation, the purple dots represent the current
positions of the Cepheids, and the red dots represent their
positions in the past.
  }
 \label{f-XY}
 \end{center}
 \end{figure*}

\section{RESULTS AND DISCUSSION}\label{results}
\subsection{Galaxy Rotation}
The system of conditional equations~(\ref{EQ-1})--(\ref{EQ-3}) has
been solved by the least squares method with weights of the form
inversely proportional to random velocity errors, with discarding
residuals by the criterion of three sigma.

As can be seen from the Fig.~\ref{f1-rot-3456}, with an increase
in the delineable unknowns, the confidence region significantly
expands with increasing $R$. The rotation curve is close to flat
one, which is in good agreement with the conclusion of
\cite{Mroz19}. Note that in the Catalog of \cite{Mroz19} there is
a ``flag'' parameter indicating whether the star was used in
kinematic analysis. The criteria of \cite{Mroz19} are more
stringent as compared to~(\ref{cut}), since it leaves only 773
stars in the sample. Our sample contains about 800 Cepheids with
estimates of their age. After applying all the rejection criteria,
our final sample contains 788 Cepheids.

Of the four cases presented in the figure, it is better to choose
the option in which the rotation curve is closest to flat one in
order to provide the most accurate spectral analysis of the
residual velocities with a minimum of false waves. So in
Fig.~\ref{f1-rot-3456}(a) and the curve goes up too early (at
$R\sim13$~kpc). The curve in Fig.~\ref{f1-rot-3456}(d) seems to
better fit the data around 12-13 kpc, but indeed with a wider
confidence interval and unnecessary wriggles at large
Galactocentric distances. As a result, the rotation curves shown
in Fig.~\ref{f1-rot-3456} ~(b) and Fig.~\ref{f1-rot-3456}~(c) can
be used to obtain the residual Cepheid velocities $\Delta
V_{circ}$ for the purpose of their spectral analysis.

For the entire sample of 788 Cepheids the following kinematic parameters were found:
 \begin{equation}
 \label{solution-all}
 \begin{array}{cll}
 (U_\odot,V_\odot,W_\odot)=\\
 (10.1,13.6,7.0)\pm(0.5,0.6,0.4)~\hbox{km s$^{-1}$},\\
      \Omega_0 =~29.05\pm0.15~\hbox{km s$^{-1}$ kpc$^{-1}$},\\
  \Omega^{'}_0 =-3.789\pm0.045~\hbox{km s$^{-1}$ kpc$^{-2}$},\\
 \Omega^{''}_0 =~0.722\pm0.027~\hbox{km s$^{-1}$ kpc$^{-3}$},\\
 \Omega^{'''}_0=-0.087\pm0.007~\hbox{km s$^{-1}$ kpc$^{-4}$},\\
           R_0 =~~8.27\pm0.10~\hbox{kpc}
 \end{array}
 \end{equation}
where the error of the unit of weight $\sigma_0=12.4$~km s$^{-1}$,
the Galaxy rotation velocity $V_0=240.2\pm3.2$~km s$^{-1}$. Note
that the solution (\ref{solution-all}) was obtained in such a way
that $R_0$ was also considered the unknown variable. The rotation
curve with parameters~(\ref{solution-all}) is shown in
Fig.~\ref{f1-2rot}.

Based on a sample of 147 masers with trigonometric parallaxes,
\cite{Reid19} found the following values of the two most important
kinematic parameters: $R_0=8.15\pm0.15$ kpc and
 $\Omega_\odot=30.32\pm0.27$ km s$^{-1}$ kpc$^{-1}$, where $\Omega_\odot=\Omega_0+V_\odot/R.$
The velocity $V_\odot=12.24$ km s$^{-1}$ was taken from
\cite{Schonrich10}). \cite{Reid19} used the expansion of the
linear Galactic rotation velocity into a series.

Based on a similar approach, \cite{Hirota2020} obtained the
following estimates from analysis of 99 masers that were observed
within the VERA program:
 $R_0=7.92\pm0.16$~(stat.)$\pm0.3$~(syst.) kpc and
  $\Omega_\odot=30.17\pm0.27$~(stat.)$\pm$0.3~(syst.) km s$^{-1}$ kpc$^{-1}$,
where $\Omega_\odot=\Omega_0+V_\odot/R,$ and the velocity
$V_\odot=12.24$ km s$^{-1}$ was also taken from
\cite{Schonrich10}.

Based on 239 Galactic masers with measured trigonometric parallaxes, \cite{Bobylev2020}
found the solar velocity components $(U_\odot,V_\odot,W_\odot)=(7.79,15.04,8.57)\pm(1.25,1.25,1.21)$ km s$^{-1}$ and the following parameters of the Galactic rotation curve:
 $\Omega_0=29.01\pm0.33$ km s$^{-1}$ kpc$^{-1}$,
 $\Omega^{'}_0=-3.901\pm0.069$ km s$^{-1}$ kpc$^{-2}$,
 $\Omega^{''}_0=0.831\pm0.032$ km s$^{-1}$ kpc$^{-3}$,
and $V_0=236.4\pm4.4$ km s$^{-1}$ for the value of $R_0=8.15\pm0.12$ kpc found.

Using a sample of 773 Classical Cepheids with precise distances
coupled with proper motions and line-of-sight velocities from Gaia
DR2, \cite{Mroz19} constructed the rotation curve of the Milky Way
up to the distance of $R\sim20$ kpc. These authors found the
rotation velocity of the Sun $V_0=233.6\pm2.8$ km s$^{-1}$ for
adopted $R_0=8.122\pm0.031$~kpc. It should be noted that the
rotation velocity $V_0$ found by us (\ref{solution-all}) is in
very good agreement with the result of the work \cite{Mroz19},
obtained from the analysis of practically the same stars.

In a recent work by \cite{Ablimit20}, around 3,500 classical
Cepheids from various sources, including \cite{Mroz19} and
\cite{Skowron19}, were used to construct the rotation curve of the
Galaxy over the distance interval $R=4-19$ kpc. The circular
rotation velocity of the solar neighborhood was obtained equal to
$V_0=232.5\pm0.9$~km s$^{-1}$ (for adopted
$R_0=8.122\pm0.031$~kpc), which is in good agreement with our
estimate.

Further, we assume that the true $r_t$ and the adopted distance $r$ are related as
$r_t=r/p,$ where $p$ is the distance-scale correction factor.
The value of the coefficient $p$ is determined
by the internal agreement of the data. Namely, by the agreement of
the line-of-sight and tangential velocities. There are two ways to
search for the value of the coefficient $p$: either by solving the
basic kinematic equations (4)--(6) where it will act as an
unknown~\citep{Rastorguev17}, or by comparing the values of the
first derivative $\Omega'_0$ obtained only from the analysis of
line-of-sight velocities, $\Omega'_0(r)$, and only tangential
velocities, $\Omega'_0(\mu)$, than $p=\Omega'_0(\mu)/\Omega'_0(r)$
\citep{Zabolotskikh02}.

We defined the value of the factor $p$ both for the entire
sample and for subsamples of different ages. As a result, we found
that the coefficient $p$ always has approximately the same value,
equal to $\sim0.9$. On this basis, it is concluded that the
distances $r$ of the analyzed Cepheids, calculated on the basis of
the period-luminosity relation, must be extended by about 10\%.

 The results of determining the $R_0$ and $V_0$ by various authors are given in Fig.~\ref{f-R0V0}, where the results are marked with the following numbers:
 (1)~-- \cite{Vallee17a}, (2)~-- \cite{Grijs17}, (3)~-- \cite{Camarillo18},
 (4)~-- \cite{Do19}, (5)~-- \cite{Mroz19}, (6)~-- \cite{Abuter19},
 (7)~-- \cite{Reid19}, (8)~-- \cite{Ablimit20}, (9)~-- \cite{Hirota2020},
 (10)~-- \cite{Bobylev2020},
 (12)~-- \cite{Majaess18}, (13)~-- \cite{Griv20},  (11)~-- this work,.

\begin{table*}\caption[]{\small
The parameters of the spiral density wave found from samples of Cepheids from three age intervals for the present moment of time
  }
\begin{center}      \label{t1}
\begin{tabular}{|l|r|r|r|r|r|}\hline
 Parameters              & $t<90$ Myr & $t:90-120$ Myr &  $t>120$ Myr \\\hline
      $\lambda_R,$ kpc  &  $  2.5\pm0.3$ & $  3.0\pm0.6$ & $ 5.1\pm1.1$ \\
            $f_R,$ km s$^{-1}$ &  $ 12.0\pm2.3$ & $  9.2\pm2.5$ & $ 6.5\pm1.5$ \\
            $i_R,$ deg &  $-10.8\pm3.1$ & $-13.1\pm3.5$ & $ -21\pm4$   \\
 $(\chi_\odot)_R,$ deg &   $  26\pm11 $ & $   52\pm10$  & $  -8\pm4$   \\
  \hline
      $\lambda_\theta,$ kpc  & $  2.7\pm0.5$ & $  2.6\pm0.7$ & $ 4.8\pm1.4$ \\
            $f_\theta,$ km s$^{-1}$ & $  8.9\pm2.5$ & $  9.6\pm2.7$ & $ 7.5\pm1.5$ \\
            $i_\theta,$ deg & $-11.8\pm3.1$ & $-11.6\pm3.8$ & $ -20\pm5$   \\
 $(\chi_\odot)_\theta,$ deg & $   58\pm12 $ & $  -51\pm12$  & $  12\pm6$   \\
  \hline
\end{tabular}
\end{center}
 \end{table*}
\begin{table*}
\caption[]{\small
Parameters of a spiral density wave found from samples of Cepheids from four age intervals in the past
  }
\begin{center}      \label{t2}
\begin{tabular}{|l|r|r|r|r|r|}\hline
 Parameters              & $t<90$ Myr & $t:90-120$ Myr &  $t>120$ Myr &  Whole sample \\\hline
      $\lambda_R,$ kpc  & $  2.6\pm0.5$ & $  2.4\pm0.8$ & $  2.7\pm0.8$ & $  2.3\pm0.4$ \\
            $f_R,$ km s$^{-1}$ & $ 12.9\pm2.6$ & $ 13.2\pm3.0$ & $  7.9\pm3.5$ & $  9.0\pm2.1$ \\
            $i_R,$ deg & $-11.6\pm3.4$ & $-10.5\pm3.6$ & $-11.8\pm3.3$ & $-10.0\pm2.4$ \\
 $(\chi_\odot)_R,$ deg & $  -74\pm15 $ & $  -44\pm17$  & $  -73\pm18$  & $  -50\pm10$  \\
  \hline
      $\lambda_\theta,$ kpc  & $  2.9\pm0.6$ & $  2.4\pm0.7$ & $  2.7\pm0.8$ & $  2.7\pm0.5$ \\
            $f_\theta,$ km s$^{-1}$ & $  8.3\pm2.6$ & $ 11.7\pm3.1$ & $  5.0\pm3.2$ & $  5.9\pm2.4$ \\
            $i_\theta,$ deg & $-12.7\pm3.5$ & $-10.5\pm3.6$ & $-11.8\pm3.3$ & $-11.9\pm2.5$ \\
 $(\chi_\odot)_\theta,$ deg & $   60\pm16 $ & $   60\pm18$  & $  -33\pm14$  & $  -81\pm12$  \\
  \hline
\end{tabular}
\end{center}
\end{table*}

 \begin{figure*}{\begin{center}
 \includegraphics[width=120mm]{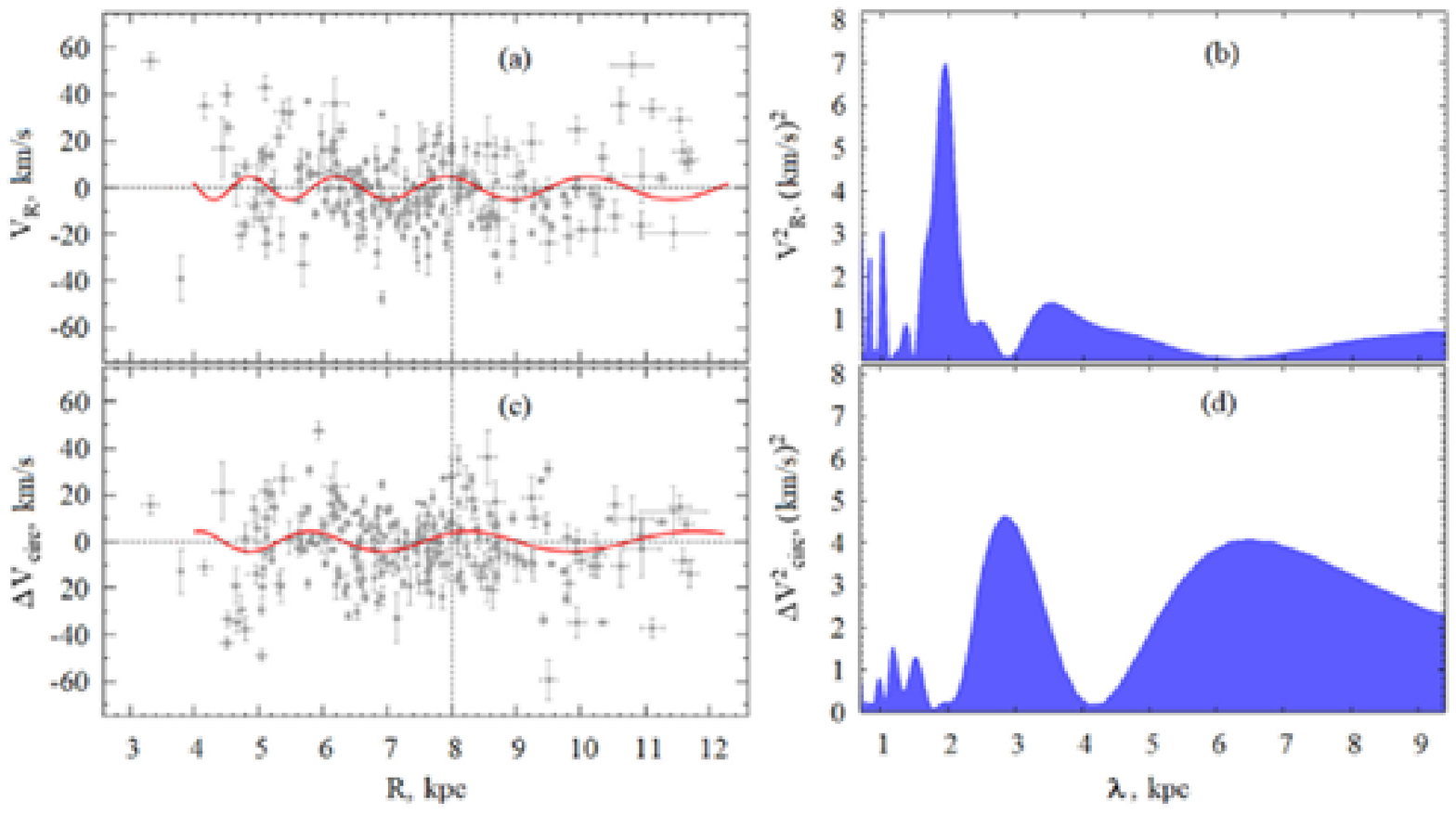}
 \caption{
In the left-hand panel radial $V_R$ (a) and residual tangential $\Delta V_{circ}$ (c) velocities of young ($t\leq90$~Myr) Cepheids are shown.  The velocities are given with error bars, the continuous periodic curves corresponding to the peaks of the power spectra (the spiral density wave) are given in red. The vertical dotted line marks the position of the Sun. In the right-hand panel the corresponding power spectra  (b) and (d) are shown.
 }
 \label{f-spectr-young}
 \end{center} }
 \end{figure*}
 \begin{figure*}
 {\begin{center}
 \includegraphics[width=120mm]{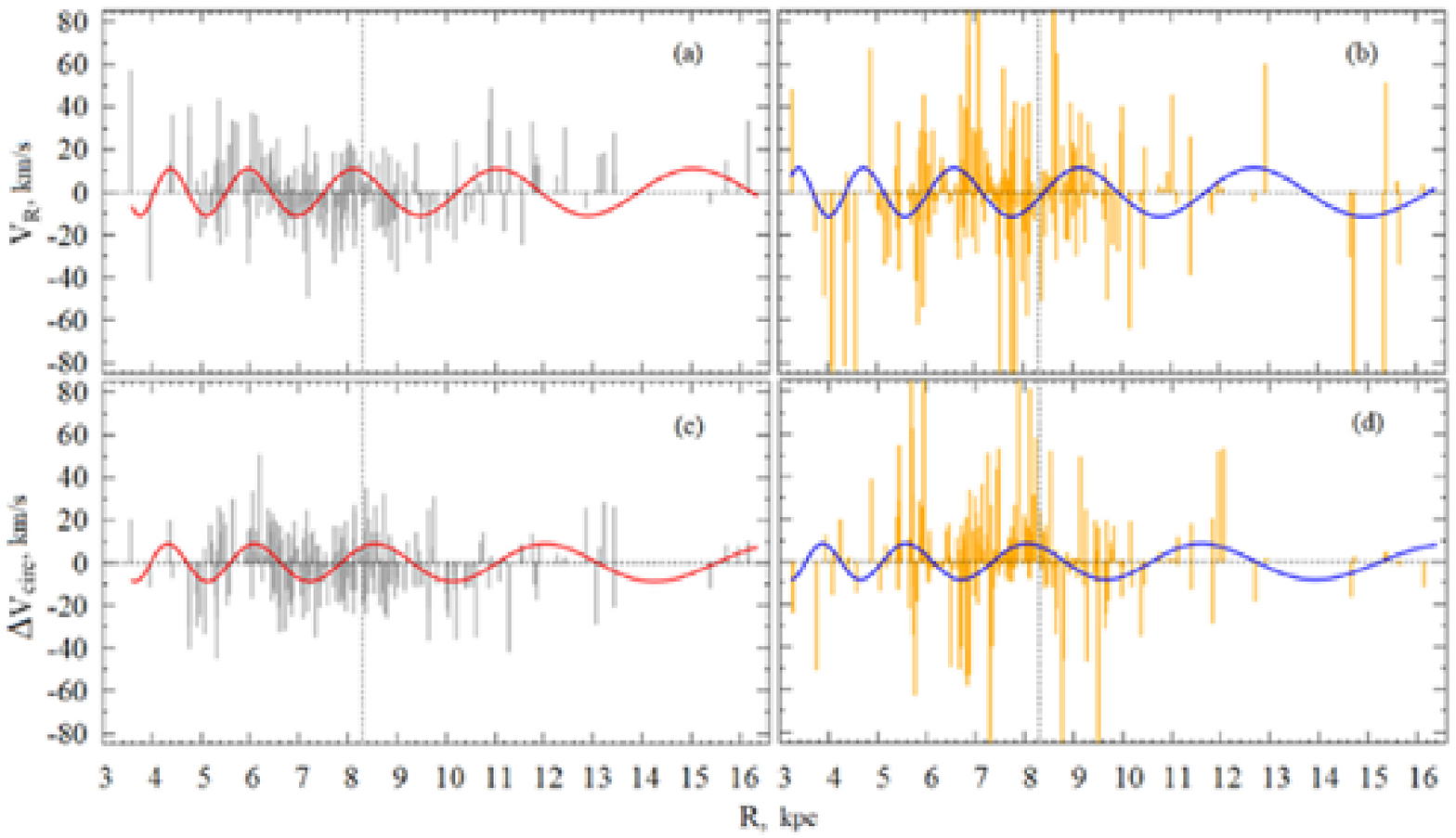}
 \caption{
The radial $V_R$ velocities of young ($t\leq90$~Myr) Cepheids at the present (a) and in the past (b), their residual tangential  $\Delta V_{circ}$ velocities at the present (c) and in the past (d). The continuous periodic curves corresponding to the peaks of the power spectra (the spiral density wave) are shown in red (in the present) and blue (in the past). The vertical dotted line marks the position of the Sun.
 }
 \label{f-VRT-young}
 \end{center} }
 \end{figure*}
 \begin{figure*}
 {\begin{center}
 \includegraphics[width=120mm]{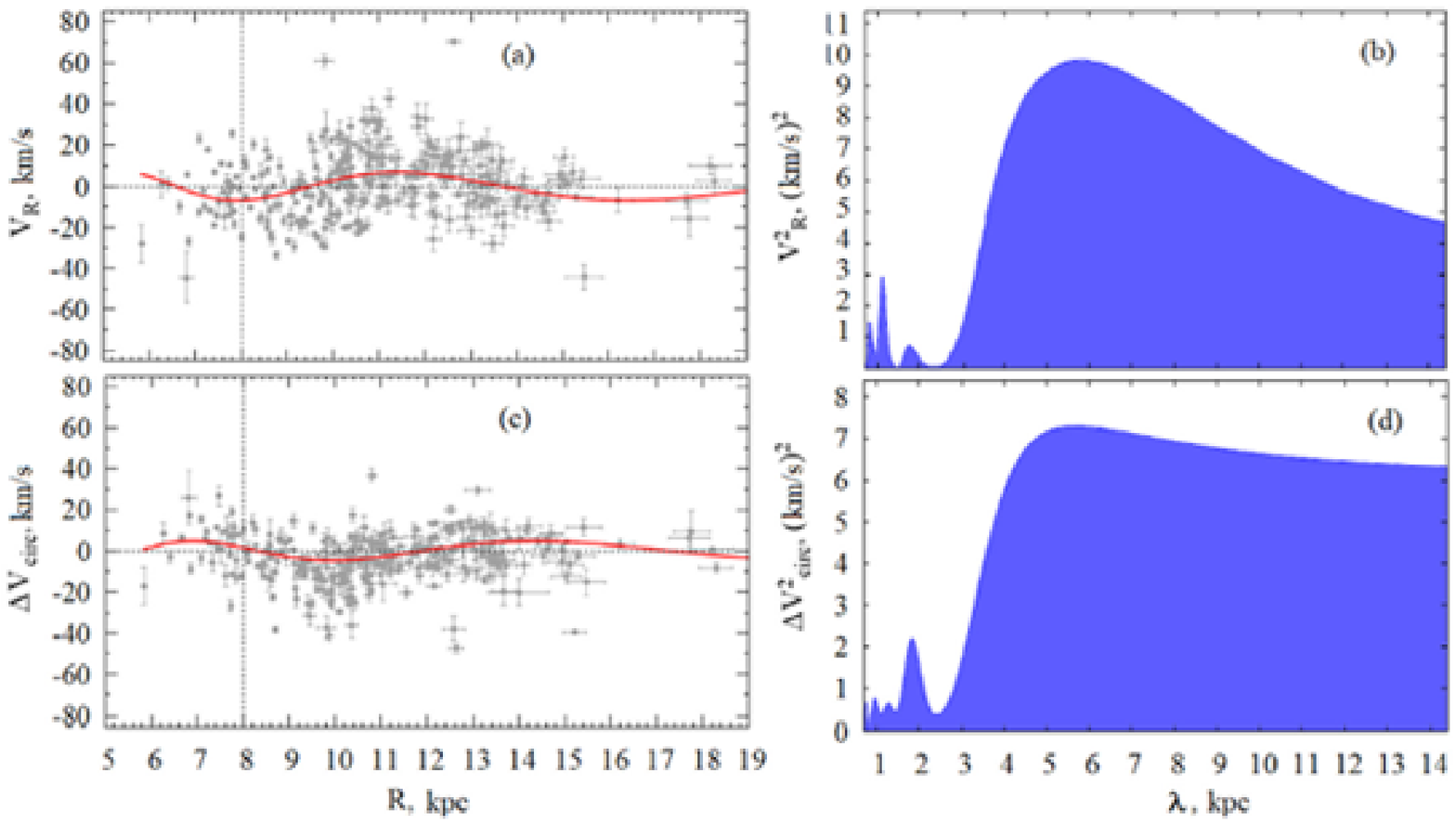}
 \caption{
In the left-hand panel radial $V_R$ (a) and residual tangential $\Delta V_{circ}$ (c) velocities of old ($t>120$~Myr) Cepheids are shown. The velocities are given with error bars, the continuous periodic curves corresponding to the peaks of the power spectra (the spiral density wave) are given in red. The vertical dotted line marks the position of the Sun. In the right-hand panel the corresponding power spectra (b) and (d) are shown.
 }
 \label{f-spectr-old}
 \end{center}}
 \end{figure*}
 \begin{figure*}
 {\begin{center}
 \includegraphics[width=120mm]{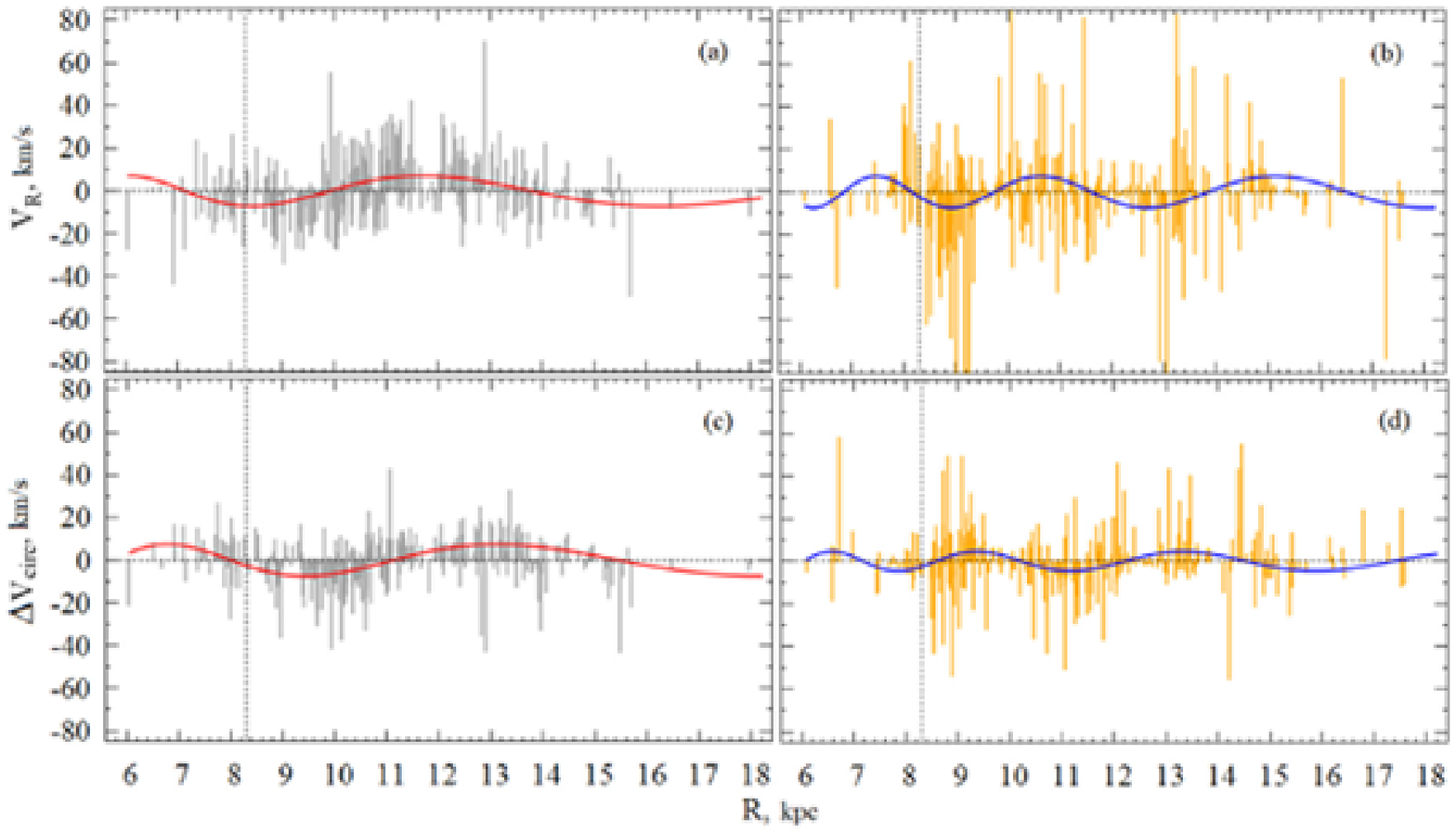}
 \caption{
The radial $V_R$ speeds of the old ($t>120$~Myr) Cepheids at the present (a) and in the past (b), their residual tangential $\Delta V_{circ}$ speeds at the present (c) and in the past (d). The continuous periodic curves corresponding to the peaks of the power spectra (the spiral density wave) are shown in red (in the present) and blue (in the past). The vertical dotted line marks the position of the Sun.
 }
 \label{f-VRT-old}
 \end{center} }
 \end{figure*}

Fig.~\ref{f1-2rot} gives two rotation curves. One corresponds to solution~(\ref{solution-all}). The second curve corresponds to the axisymmetric gravitational potential model~III (a modified NFW model) from the work of \cite{Bajkova16}:
 \begin{equation}
 \renewcommand{\arraystretch}{2.8}
 \begin{array}{cll}\displaystyle
  V^2_{circ}=\frac{\displaystyle M_b R^2}{\displaystyle (R^2+b_b^2)^{3/2}}
            +\frac{\displaystyle M_d R^2}{\displaystyle [R^2+(a_d+b_d)^2]^{3/2}}\\
   +M_h \biggl[\frac{\displaystyle \ln(1+R/a_h)}{\displaystyle R}
   -\frac{\displaystyle 1}{\displaystyle R+a_h}\biggr]
   +const,
 \label{Vcirc-III}
 \end{array}
 \end{equation}
where, $M_b, M_d$ and $M_h$ are the masses of the bulge, disk and halo respectively, $b_b, a_d, b_d$ and $a_h$ are the scale lengths (in kpc) of the corresponding galactic components. The gravitational potential is expressed in units of 100 km$^2$~s$^{-2}$, the lengths in kpc, and the masses in galactic mass units $M_g=2.325\times 10^7 M_\odot$ providing the value of the gravitational constant $G=1.$ The term $const$ is needed here to accurately match the solar rotation velocity $V_0$ in this work and in the work of \cite{Bajkova16} ($const=-4.8$~km$^2$ s$^{-2}$).

To construct a curve, it is necessary to substitute the following
values of seven parameters into this formula~(\ref{Vcirc-III}):
 $M_b= 44300~(M_g),$
 $b_b=0.2672$~kpc,
 $M_d=279800~(M_g),$~
 $a_d=  4.40$~kpc,
 $b_d=0.3084$~kpc,
 $M_h=1247400~(M_g),$~
 $a_h= 7.7$~kpc.

For spectral analysis of residual circular velocities $\Delta
V_{circ}$ it is important that they are obtained with a relatively
smooth rotation curve. As can be seen from Fig.~\ref{f1-2rot}, the
curve~(\ref{Vcirc-III}) can be used to obtain residual circular
velocities in a very wide range of distances $R>4$~kpc. The
applicability of the rotation curve corresponding to the
solution~(\ref{solution-all}) is limited by the interval
$R:4-20$~kpc. In spectral analysis, we use both of the rotation
curves described above for mutual control.

 \begin{figure*}{\begin{center}
 \includegraphics[width=140mm]{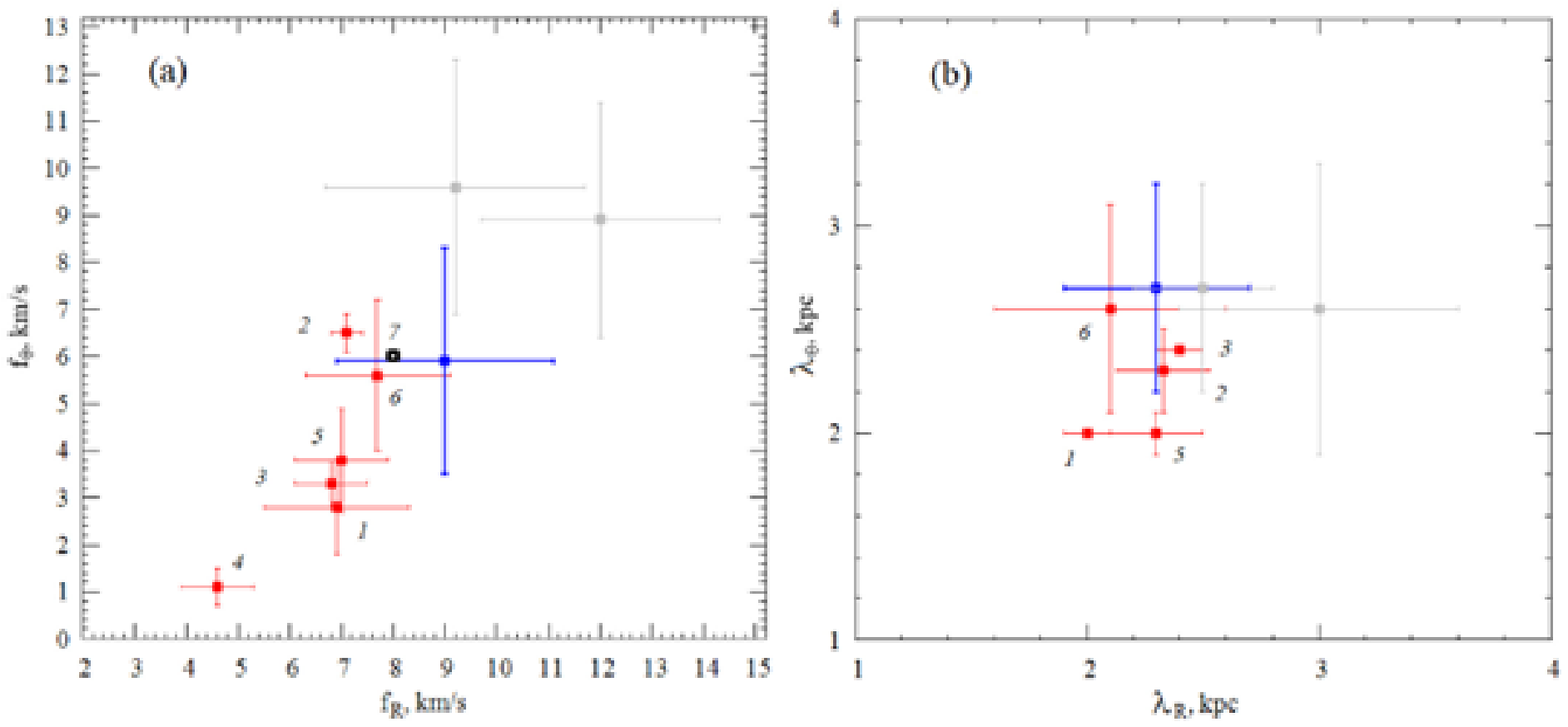}
 \caption{
The results of determining the velocities $f_R$ and $f_\theta$
~(a) and  velocity $\lambda_R$ and $\lambda_\theta$ (~b) by
various authors -- red and black, blue and gray --- the result of
this work, see also the text.}
 \label{f-fR-lambda}
 \end{center}
  }
 \end{figure*}

 \subsection{Spiral Density Wave Parameters}
Spectral analysis was performed for Cepheids of three samples of
different ages. The age  boundaries were chosen so to ensure
approximately equal number of stars in the samples. For each sample,
the spiral density wave parameters were obtained for two cases.
In the first case, spectral analysis was performed
for the present moment in time. In the second case, the position
and speeds of each Cepheid were taken at the time of their birth.
That is, a galactic orbit in the past was constructed for each
star in accordance with an estimate of its age. To construct
galactic orbits in the past, an axisymmetric model of the
gravitational potential of the Galaxy was used (a modified NFW
model from \cite{Bajkova16}).

The first sample contains 254 stars satisfying the condition of
$t\leq90$~Myr. The average age of these relatively young Cepheids
is ${\overline t}=66$~Myr.

The second sample contains 249 stars with ages from the interval
90--120~Myr The average age of these Cepheids is
${\overline t}=105$~Myr.

The third sample contains 304 stars selected under the condition
$t>120$~Myr. The average age of these Cepheids is ${\overline
t}=165$~Myr.

Fig.~\ref{f-XY} shows the $X,Y$ distribution
of three  samples of Cepheids at the present time and in the past
according to the age of each star. Parameters given on
Fig.~\ref{f-XY} of a four-arm spiral pattern were found by
\cite{BobBaj14} from masers with measured trigonometric
parallaxes. In the Figure, the following segments of spiral arms
are numbered in Roman numerals: I~--- the Scutum arm, II~--- the
Carina--- Sagittarius arm, III~--- the Perseus arm, IV~--- the
Outer arm.

The results of the spectral analysis of Cepheids are reflected in
the Tables~\ref{t1}--\ref{t2}, as well as on
Figures~\ref{f-spectr-young}--\ref{f-VRT-old}.

The Table~\ref{t1} gives the parameters of the spiral density wave
found from samples of Cepheids from three age intervals for the
current moment of time. In the Table~\ref{t2} similar values are
given for four samples of Cepheids calculated for their past
positions. In the Table~\ref{t2} a column with the results
obtained for the entire sample is added.

One of the most important parameters determining on the basis of
spectral analysis is the wavelength $\lambda$. With the found
value $\lambda$, the pitch angle $i$ is calculated using the
relation~(\ref{a-04}). As can be seen from the first and second
columns of the Tables~\ref{t1}--\ref{t2}, for samples of Cepheids
younger than 120 Myr, the values of $\lambda$ lie in the range
2.4--3.0 kpc (this means that $i$ is in the range  [$-13^\circ$
$-10^\circ$] for a four-arm pattern model, $m=4$).

The values of $\lambda_R$ and $\lambda_\theta$, found from a
sample of Cepheids older than 120 Myr for the current moment of
time (Table~\ref{t1}), are very different from the similar values
found from younger Cepheids. This problem is eliminated only in
the case of an analysis of the velocities of the old Cepheids
calculated at the time of their birth (Table~\ref{t2} and
Fig.~\ref{f-spectr-old}--\ref{f-VRT-old}).

Fig.~\ref{f-spectr-young} shows the radial $V_R$ and residual
tangential $\Delta V_{circ}$ velocities at the present obtained
for a sample of young ($t\leq90$~Myr) Cepheid and their spower
spectra. For the same Cepheids, Fig.~\ref{f-VRT-young} gives the
radial $V_R$ and residual tangential $\Delta V_{circ}$ velocities
versus the distance $R$ at the present time and in the past, where
the periodic curves show the effect of a spiral density wave. The
first column of the Table~\ref{t1} gives the values of the spiral
wave parameters found using the young ($t\leq90$~Myr) Cepheids. It
should be noted that both for the sample of young Cepheids and for
Cepheids of intermediate age, the values of the parameters of the
spiral wave indicated in both tables are close. Therefore,
illustrations for Cepheids of intermediate age are not given.

In Fig.~\ref{f-spectr-old}, for the sample of old ($t>120$~Myr)
Cepheids, radial $V_R$ and residual tangential $\Delta V_{circ}$
velocities are given at the present and their power spectra. The
radial $V_R$ and residual tangential $\Delta V_{circ}$ velocities
are given in Fig.~\ref{f-VRT-old} at the present time and in the
past.

Another important parameter, to be determined on the basis of
spectral analysis, is the amplitude of the perturbations $f_R$ or
$f_\theta $. If we take the peak values of the squares of
velocities from the power spectra in Fig.~\ref{f-spectr-young} or
Fig.~\ref{f-spectr-old}, then the values $f_R$ or $f_\theta$
(indicated in the tables) can be found by the
formula~(\ref{Speak}).

An analysis of modern data shows that in a wide region of the
solar neighbourhood, the velocities $f_R$ and $f_\theta$ are
usually 4--9~km s$^{-1}$, and the wavelength $\lambda $ is in the
range 2--3~kpc.

Thus, from 130 maser sources with measured trigonometric
parallaxes in the work of \cite{Rastorguev17}, there were found
$f_R=6.9\pm1.4$~km s$^{-1}$ and $f_\theta=2.8\pm1.0$~km s$^{-1}$,
solar phase $\chi_\odot=-125^\circ\pm10^\circ$. From 239 Galactic
masers with measured trigonometric parallaxes in the work of
\cite{Bobylev2020}, there were found $f_R=7.0\pm0.9$~km s$^{-1}$
and $f_\theta=3.8\pm1.1$~km s$^{-1}$.

From about 500 OB stars of the Gaia DR2 Catalog, \cite{BobylevBaj18} determined
 $f_R=7.1\pm0.3$~km s$^{-1}$ and $f_\theta=6.5\pm0.4$~km s$^{-1}$,
 $\lambda_R=2.3\pm0.2$~kpc and $\lambda_\theta=2.3\pm0.2$~kpc,
$(\chi_\odot)_R=-135^\circ\pm5^\circ$ and
$(\chi_\odot)_\theta=-123^\circ\pm8^\circ$. From sample of open
clusters younger than 50 Myr, \cite{BobShirokova16} determined
$f_R=7.7\pm1.4$~km s$^{-1}$ and $f_\theta=5.6\pm1.6$~km s$^{-1}$,
$\lambda_R=2.1\pm0.5$~kpc and $\lambda_\theta=2.6\pm0.5$~kpc,
$(\chi_\odot)_R=-85^\circ\pm10^\circ$ and
$(\chi_\odot)_\theta=-62^\circ\pm9^\circ$.

From about 200 Cepheids from the Hipparcos Catalog,
\cite{BobylevBaj12} found $f_R=6.8\pm0.7$~km s$^{-1}$ and
$f_\theta=3.3\pm0.5$~km s$^{-1}$, $\lambda=2.0\pm0.1$~kpc,
$\chi_\odot=-193^\circ\pm5^\circ$. We also note the new values of
$f_R=4.6\pm0.7$~km s$^{-1}$ and $f_\theta=1.1\pm0.4$~km s$^{-1}$,
obtained in a recent work by \cite{Loktin19} from the analysis of
modern data on Open Star Clusters.

Note that~\citep{Burton1971} calculated the expected values for the perturbation velocities
 $f_R\approx8$~km s$^{-1}$ and $f_t\approx6$~km s$^{-1}$ for $R=8$~kpc.

The results of determining the $f_R,$ $f_\theta$ and $\lambda_R,$
$\lambda_\theta$ by various authors are given in
Fig.~\ref{f-fR-lambda}, where the results are marked with the
following numbers: (1)~-- \cite{Rastorguev17}, (2)~--
\cite{BobylevBaj18}, (3)~-- \cite{BobylevBaj12}, (4)~--
\cite{Loktin19}, (5)~-- \cite{Bobylev2020},  (6)~--
\cite{BobShirokova16}, (7) and black square~-- \cite{Burton1971};
results of this work from the last column of the table~\ref{t2}
are shown in blue, from first and second column of the
table~\ref{t1} are shown in gray.

Currently, there is no single generally accepted model of the spiral structure of the Galaxy. Theorists usually use the simplest two-arm model with a pitch angle of 5--7 degrees. Modern data on the distribution of clouds of neutral hydrogen, ionized hydrogen,
and maser sources speak rather of a four-arm model with a pitch
angle of 10--14 degrees. Reviews on this issue can be found, for
example, in the works \cite{HouHan2014} or \cite{Vallee2017c}. In
this paper, we adhere to the four-arm spiral pattern model with
the parameters we found in previous works from the spatial and
kinematical analysis of maser sources \citep{BobBaj14}.

In work of Dambis et al. (2015), an analysis of the spatial distribution of a large sample of classical Cepheids yielded estimates of the pitch angle of the four-arm spiral pattern
$i=-9.5^\circ\pm0.1^\circ$ and the solar phase $\chi_\odot=-121^\circ\pm3^\circ$.

The model of the global four-arm spiral pattern in the Galaxy is
defended, for example, in the works of \cite{Vallee17b,Vallee18}.
Some authors prefer so far (before the appearance of more
high-precision data) to consider individual segments of spiral
arms with individual pitch angles \citep{Nikiforov18}. In these
models, the pitch angle lies in the range from $-10^\circ$ to
$-15^\circ$.

Light absorption strongly distorts the spatial distribution of
objects and makes it difficult not only to determine their
photometric distances, but also to calibrate the PLR even by
high-precision trigonometric distances. To determine the absolute
magnitude, it is necessary to know the color excesses, and for the
Cepheids of the Milky Way this is a much more serious problem than
for stars of constant brightness \citep{Rastorguev13,Lazovik20} or
for LMC Cepheids. Supposing a unified absorption law, we can solve
these problems and reduce the effect of differential absorption by
using two-color Wesenheit indices ($W_{VI}$ as an example) instead
of absolute magnitudes \citep{Madore1976}. To use Wesenheit
indexes of Cepheids instead of PLR, we should know the
``period--normal color'' relation, which reflects the shape and
width of the instability strip and strongly depends on the
metallicity of young populations. In addition, the use of
Wesenheit indices or tricolor indices like $Q_{UBV}$ to reduce the
effects of differential absorption, is justified only for a small
optical depth of dust in the broadband (heterochromic) photometry,
which does not take place for Milky Way disk in optics.

The problem of calibrating PLRs and estimating photometric
distances is additionally complicated by a noticeable difference
in the absorption laws in the Milky Way galaxy
\citep{Fitzpatrick07, Fitzpatrick09}, differing, first of all, by
the value $R_V=A_V/E_{B-V}$, which varies from 2 to 6. That is why
the photometric distances of the objects in the galactic disk,
determined by optical data, can be strongly suffered from both
random and hard-to-account for systematic errors.

It is for these reasons that in recent years, photometric data in NIR/MIR (2MASS, AllWIZE, Spitzer and other projects) have been used to determine the photometric distances of objects, including Cepheids. For example, light absorption (expressed in magnitudes)
in the $K_s$ (2MASS) band $A_{K_s}\approx 0.078 (\pm0.004)\cdot Av$, and in the WISE $W1$, $W2$ bands is $A_{W1}\approx 0.039 (\pm0.004) \cdot Av$ and $A_{W2}\approx 0.026 (\pm0.004)\cdot Av$ respectively \cite{Wang19}. As is well known, average specific
absorption $a_V$ in the Galaxy plane is about 1.5 mag kpc$^{-1}$
and, as a result, up to distances of 5 kpc in the $W2$ band the
absorption not exceed 0.20 mag (which is comparable with the
internal scatter of PLR), and the effect of differences in the
absorption laws is negligible at all. It should also be noted that
the study of structures in the Galaxy, such as a spiral pattern,
based on the spatial distribution of objects, is greatly affected
by the selection effects due to interstellar extinction even in
IR. However, these selection effects do not affect the kinematics
at all, which justifies the kinematic analysis of the spiral
structure performed in this paper.

As for microlensing effects, it should be borne in mind that for bulge stars this effect should be taken into account due to the huge number of observed stars (this is exactly what the results of the OGLE project discovered). The total sample size of Cepheids is
negligible compared to the sample of bulge stars studied in OGLE,
and even with the same lensing probability, the average expected
number of lensings for the entire sample is much less than unity.
Also, the lensing event itself, lasting some tens of days, in
principle, could not distort either the estimate of the period
(since the photometric monitoring of Cepheids lasts much longer
than this time), or the results of precision astrometric
measurements by Gaia, conducted for about 30 months.

Most recently, the latest version of the Gaia Early Data Release 3 \citep{Brown20, Lindegren20} Catalog was published. It clarifies by about 30\% the values of trigonometric parallaxes and proper motions for about 1.5 billion stars. The radial velocities are
simply copied from the Gaia DR2 Catalog. We hope that the use of new data will not have a fundamental impact on the conclusions of this work.

Flat rotation curves of young objects reaching distances of about 15--20 kpc from the Galactic center, derived in numerous papers cited here, show an almost linear increase in the effective mass with the distance. A simple estimate of the effective mass within
a radius of 20 kpc for a rotation velocity of 220--230 km s$^{-1}$ leads to ($2.3\pm0.3)\cdot 10^{11}$ solar masses. The main contribution even to this mass gives the dark matter, whose
contribution only dominates with a further increase in distance.
Its total contribution to the mass of the Galaxy and the local
density of gravitating matter can be estimated only by modelling
the gravitational potential, which takes into account the
contribution of all structural components of the Galaxy to the
rotation curve. In particular, this was done in the cited papers
of \cite{Ablimit20}) and \cite{Bajkova16}. In the latter paper the
parameters of the density laws were derived from the kinematical
study not only of disk objects, but also of very distant halo
objects, and therefore they are considered as more reliable as
compared to ones based only on objects within 15-20 kpc from the
Galaxy center.

In this paper it was shown that the rotation curve of Cepheids is well approximated by the theoretical three-component model III from \cite{Bajkova16}. Taking the estimates of the parameters of NFW model and the appropriate errors, we can easily derive an
estimate of the contribution of DM to the total local density of gravitating matter:
 $\rho_{DM}\approx0.0114^{+0.0078}_{-0.0049}~M_\odot\cdot {\rm pc}^{-3}$, which is 5
to 20\% of the contribution of baryonic matter ($0.101 M_\odot\cdot {\rm pc}^{-3}$) according to the latest estimate made by the authors of the Besancon model of the Milky Way
\citep{Mor18}, based on the entire set of observational data. Our
estimate of DM's contribution, in particular, is in excellent
agreement with those of
 \citep{Mor18} ($0.012\pm0.001 M_\odot\cdot {\rm pc}^{-3}$);
 \cite{de Salas19} ($0.008-0.011 M_\odot\cdot {\rm pc}^{-3}$) and
 \cite{Ablimit20} ($0.0105\pm0.0012 M_\odot\cdot {\rm pc}^{-3}$).
Note also that all the above estimates agree with an upper limit
on the DM contribution to the local density in the Milky Way
galaxy ($0.027 M_\odot\cdot {\rm pc}^{-3}$), made on the basis of
a completely different approach -- the analysis of
galacto-vertical oscillations of Cepheids \citep{Dambis04} and
young open clusters \citep{Dambis03}.

\section{CONCLUSIONS}
The spatial and kinematic properties of a large sample of classical Cepheids with proper motions and line-of-sight velocities from the Gaia DR2 Catalog were studied. For this, we
used data from the works of \cite{Mroz19} and \cite{Skowron19}. The final sample contains about 800 Cepheids. For each of them there are estimates of distance and age as well.

The parameters of galactic rotation were found over the entire
sample of Cepheids. So, the linear speed of rotation of the Galaxy
at a solar distance amounted to $V_0=240\pm3$~km s$^{-1}$.
Moreover, the distance from the Sun to the axis of rotation of the
Galaxy was found to be equal to $R_0=8.27\pm0.10$kpc. We found
that the distance scale correction factor $p$ for both the entire
sample and sub-samples of different ages has approximately the
same value, equal to $\sim0.9$. On this basis, it is concluded
that the distances $r$ of the analyzed Cepheids, calculated on the
basis of the period-luminosity relation, must be extended by about
10\%.

There was performed a spectral analysis of both radial $V_R$, and
residual tangential velocities $\Delta V_{circ}$ of Cepheid
samples of different ages. For each sample, the parameters of the
spiral density wave were obtained for two cases. In the first
case, spectral analysis was performed for the present moment of
time. In the second case, the position and speed of the Cepheids
were taken at the time of their birth. That is, a galactic orbit
in the past was constructed for each star in accordance with an
estimate of its age.

A spectral analysis of radial and tangential velocities showed that for samples of Cepheids younger than 120 Myr, both at the present time and in the past, we obtain close estimates of the parameters of the spiral density wave. So, the value of the
wavelength $\lambda_{R,\theta}$ lies in the range of [2.4--3.0]
kpc, the pitch angle $i_{R,\theta}$ in the range of [$-13^\circ,
-10^\circ$] for the four-arm pattern model, the amplitude of the
radial perturbations is $f_R\sim12$~km s$^{-1}$, and the
tangential perturbations are $f_\theta\sim9$~km s$^{-1}$. These
values are in agreement with the results of the analysis of other
young objects of the Galaxy (for example, maser sources or OB
stars).

But the sampling rates of older Cepheids (over 120 Myr) at the present time give the wavelength $\lambda_{R,\theta}\sim5$~kpc (hence $i\sim20^\circ$). This value contradicts the known results.
This means that a lot of time has passed since the birth of these
Cepheids in the spiral arms, they are significantly removed from their place of birth, and at present the sample does not have coherent properties.  An analysis of positions and velocities of old Cepheids (more than 120 Myr), calculated by integrating their orbits backward in time, made it possible to determine significantly more
reliable parameters of the spiral density wave: wavelength $\lambda_{R,\theta}=2.7$~kpc, amplitude of radial and tangential perturbations $f_R=7.9$~km s$^{-1}$ and $f_\theta=5$~km s$^{-1}$ respectively.

 \section*{Acknowledgments}
The authors would like to express their sincere gratitude to the anonymous referees for the interesting and useful remarks, the consideration of which made it possible to significantly improve the article. A.~Rastorguev and M.~Zabolotskikh are grateful to the
Russian Foundation of Basic Research (Grant No.~19--02--00611) for
partial financial support.

 \section*{DATA AVAILABILITY}
The data underlying this article will be shared on reasonable
request to the corresponding author.

 \end{document}